\documentclass[showpacs,showkeys,12pt,preprint,preprintnumbers,nofootinbib,groupedaddress,superscriptaddress,amsmath,amssymb]{revtex4}

\usepackage[dvips]{color} 
\usepackage{graphicx}
\usepackage{epsfig} 
\usepackage{dcolumn} 
\usepackage{bm} 
\usepackage{here}

\newcommand{\gsim}{\mbox{ \raisebox{-1.0ex}{$\stackrel{\textstyle >}
{\textstyle \sim}$ }}}
\newcommand{\lsim}{\mbox{ \raisebox{-1.0ex}{$\stackrel{\textstyle <}
{\textstyle \sim}$ }}}

\allowdisplaybreaks[4]

\begin{document}

\title{
Dimension-six top-Higgs interaction and 
its effect in collider phenomenology
}

\author{Shinya Kanemura}
\email{kanemu@sci.u-toyama.ac.jp}
\affiliation{Department of Physics, University of Toyama, 3190 Gofuku,
Toyama 930-8555, Japan}

\author{Daisuke Nomura}
\email{dnomura@pa.msu.edu}
\affiliation{Department of Physics and Astronomy, 
Michigan State University, East Lansing, Michigan 48824-2320, USA}

\author{Koji Tsumura}
\email{ko2@het.phys.sci.osaka-u.ac.jp}   
\affiliation{Department of Physics, Osaka University, Toyonaka, Osaka
 560-0043, Japan}

\preprint{UT-HET 002}
\preprint{MSUHEP-060525}
\preprint{OU-HET 562}



\begin{abstract}
 
 Measurement of the Yukawa interaction between the top quark and
 the Higgs boson should be useful to clarify the mechanism of fermion
 mass generation. 
 We discuss the impact of non-standard interactions characterized by
 dimension-six operators on the effective top Yukawa coupling.
 The cross section of the process $e^-e^+ \to W^-W^+ \nu \bar \nu \to
 t \bar t \nu \bar \nu$ is calculated including these operators,
 and possible deviation from the standard model prediction is
 evaluated under the constraint from perturbative unitarity and
 current experimental data.
 We find that if the new physics scale is in a TeV region, 
 the cross section can be significantly enhanced
 due to the non-standard interactions.
 Such a large effect should be detectable
 at the International Linear Collider.  

\end{abstract}

\maketitle

\section{Introduction}

 To establish our understanding on the origin of 
 the mass of elementary particles 
 is the top priority for today's high energy physics.
 The idea of spontaneous breakdown of the electroweak gauge symmetry
 provides a simple and compelling phenomenological picture
 for this issue in the Standard Model (SM) and its extended versions. 
 The Masses of weak gauge bosons are generated by the Higgs mechanism,
 and those of quarks and charged leptons are given via the Yukawa
 interaction after the symmetry breaking.

 In the SM a scalar iso-doublet field, the Higgs field, is introduced
 to trigger the spontaneous symmetry breaking,
 whose energy density determines the structure of vacuum.
 It is a remarkable feature of the SM 
 that all the massive elementary particles but neutrinos obtain
 masses which are proportional to the vacuum expectation value of the Higgs
 field. 
 The scale for the mass of each particle
 is determined by the magnitude of its coupling with the Higgs boson.
 In the SM, therefore,
 hierarchical structure among quark masses observed at
 experiments is a consequence of differences in the strength
 of the Yukawa interaction for each quark.
 As a result the question of the quark mass hierarchy remains
 a problem, which must be solved in the framework of
 new physics beyond the SM. 

 The top quark is exceptionally heavy as compared to the other quarks. 
 Its mass has been measured to be at the scale of
 electroweak symmetry breaking, so that its Yukawa coupling
 has turned out to be of order one within the SM.  
 This fact would indicate an important insight that the top quark 
 is deeply related to dynamics of electroweak symmetry breaking.
 There have been lots of works in this direction;
 the idea of the top mode condensation and top color models\cite{topmode,topcolor}, top flavor
 models\cite{topflavor,He:2001fz} etc. 
 These models generally predict rather strong dynamics for the
 electroweak symmetry breaking. 
 Measuring the interaction between the Higgs boson and the top quark
 is essentially important not only to confirm the SM but also
 to test new physics including these models.

 At the CERN Large Hadron Collider (LHC), which will start its operation
 in 2007, 
 it is strongly expected that the Higgs boson
 will be discovered in a wide range of its mass up to 1 TeV\cite{lhc}.
 Once the Higgs boson is found, its property such as the mass, the width, 
 production cross sections and decay branching ratios will be measured
 as precisely as possible for the purpose of testing the SM and its extensions. 
 Information of Higgs coupling constants with fermions can be 
 extracted from these observables.  
 However, measurement of the top Yukawa coupling 
 may be challenging due to the huge QCD backgrounds. 
 Precise determination of the coupling constants can be performed 
 at the International Linear Collider (ILC)\cite{ilc}. 
 At the ILC, the top Yukawa interaction is expected to be measured 
 through the process $e^-e^+ \to t \bar t H$\cite{tth} for a relatively light 
 Higgs boson when it is kinematically allowed. 
 For a heavier Higgs boson, it can be measured 
 via the vector boson fusion process 
 $e^-e^+ \to  W^-W^+\nu\bar\nu \to t \bar t \nu\bar\nu$\cite{w-fusion,Kauffman:1989aq,alcaraz}.
 
 Impact of non-standard interactions on the coupling of the top quark 
 with the Higgs boson is of central interest in this paper.
 Below the new physics scale (i.e., the cutoff scale of the SM), such a new interaction is 
 effectively described in terms of  higher dimensional operators after 
 the heavy new physics particles are integrated out.  
 In particular, the low energy effect of new physics
 is expressed at the leading order 
 by dimension-six operators in a gauge invariant Lagrangian. 
 These dimension-six operators have been systematically studied 
 in the literature\cite{buchmuller,gounaris1,gounaris2}. 
 Some of them effectively give modifications 
 to the top Yukawa coupling. 
 Han et al. have discussed the effect of the dimension-six operators 
 on the process of $e^-e^+ \to t \bar t H$\cite{han}.   

 We here calculate the cross section of the process 
 $e^-e^+ \to W^-W^+ \nu \bar \nu \to t \bar t \nu \bar \nu$ 
 by adding these dimension-six operators to the SM Lagrangian, 
 and evaluate possible deviation from the SM prediction 
 under the constraint from perturbative unitarity\cite{pwu} and
 current experimental data\cite{pdg}.
 This process  $e^-e^+ \to W^-W^+ \nu \bar \nu \to t \bar t \nu \bar \nu$ 
 has at first been studied by Yuan\cite{yuan}, Gintner and Godfrey\cite{godfrey} 
 in the SM, and its QCD correction has been studied by Godfrey and Zhu\cite{godfrey2}.  
 Larios et al. have investigated the same process in the context of 
 the SM without the Higgs boson, instead including 
 the dimension-five operators 
 $W^-W^+ \bar t t$ and $W^-_\mu W^+_\nu \bar t \sigma^{\mu\nu} t$\cite{larios}.  
 In the present paper, we keep the Higgs boson mass to be the electroweak
 scale, and classify the non-standard interaction between the Higgs boson
 and the top quark in terms of the dimension-six operators with 
 setting the cutoff scale to be TeV scales. 
 We find that the effect of the dimension-six operators with the cutoff
 scale to be a TeV region can enhance 
 the cross section significantly  especially for relatively
 large Higgs boson masses, so that the effect should be detectable
 at the ILC.  

 This paper is organized as follows. 
 In Sec. II, we introduce dimension-six operators which directly 
 affect the Higgs boson interaction with the top quark.
 Theoretical and experimental bounds for these anomalous 
 couplings are also discussed.  
 In Sec. III, we calculate the cross section of the 
 subprocess $W^-W^+ \to t \bar t$ for each helicity set of the gauge bosons.
 A test of the amplitudes by using the equivalence theorem is also
 performed. 
 In Sec. IV, we present the numerical evaluation of the deviation from the 
 SM predictions.
 Conclusions are given in Sec. V. 
 Some detailed analytic expressions are shown in 
 Appendix.
  
\section{Dimension-six operators}
\label{Sec:Dimension-six operators}

The effect of new physics can be expressed 
in terms of higher dimensional operators, 
which are induced by integrating out the 
heavy non-standard particles.
We here discuss such dimension-six operators 
which are relevant to the interaction between 
the Higgs boson and the top quark\footnote{The gauge symmetry
prohibits dimension-five operators.}. 

Below the new physics scale (i.e., the SM cutoff scale) $\Lambda$, 
the non-standard interaction can be written 
in the effective Lagrangian as
\begin{align}
{\mathcal L}^{\rm eff} =& {\mathcal L}_{\rm SM}^{}
              + {\mathcal L}_{\rm dim.6}^{}
              + {\mathcal L}_{\rm dim.8}^{} + \cdot\cdot\cdot,
\end{align}
where ${\mathcal L}_{\rm SM}^{}$ is the SM Lagrangian, and  
\begin{align}
{\mathcal L}_{{\rm dim.}n} =& 
\frac{1}{\Lambda^{n-4}} \sum_i C_i^{(n)} {\mathcal O_i^{(n)}},
 \;\;\;\; (n \geq 6), 
\end{align}
where ${\mathcal O}^{(n)}_i$
are dimension-$n$ operators which are  
$SU_C (3)\times SU_L(2) \times U_Y(1)$ invariant, 
and $C^{(n)}_i$ are the constants which represent the coupling 
strengths of ${\mathcal O}^{(n)}_i$.
Leading effects of new physics can well be described
by the dimension-six operators ${\mathcal O}^{(6)}_i$.
The higher order operators can only become important at
the scale close to  $\Lambda$. In this paper, we concentrate on 
the effect of the dimension-six operators neglecting 
additional contributions from the higher order operators
such as dimension-eight ones.

A systematic study for dimension-six operators has been 
given in the literature\cite{buchmuller}.  
The complete list of the dimension-six gauge invariant 
operators is given in Refs.~\cite{buchmuller,gounaris1}.
In this paper, we only consider the CP-conseving operators.
Then, the operators which directly modify the top-Yukawa interaction 
are the followings\cite{whisnant,FengLiMaalampi}: 
\begin{align}
{\mathcal O}_{t1} =& \left(\Phi^\dag \Phi - \frac{v^2}{2}\right) 
\left( \bar{q}_L t_R \tilde{\Phi} + \mbox{h.c.}\right),\label{eq:Ot1}\\
{\mathcal O}_{Dt} =& \left( \bar{q}_L D_\mu t_R \right) 
 D^\mu \tilde{\Phi} + \mbox{h.c.}, \label{eq:ODt}
\end{align} 
where $q_L^{}=(t_L^{},b_L^{})^T$, $\Phi$ is the scalar isospin doublet (the Higgs doublet) with
hypercharge $Y=1/2$, and   $\tilde{\Phi} \equiv i \tau_2 \Phi^\ast$
with $\tau_i$ ($i=$1-3) being the Pauli matrices.  
The doublet filed $\Phi$ is parameterized as
\begin{align}
 \Phi = \left[ \begin{array}{c} \omega^+ \\ \frac{1}{\sqrt{2}}(v+H+iz)
               \end{array}\right],
 \end{align}
 where  $\omega^\pm$ and $z$ are the Nambu-Goldstone bosons, 
 $H$ is the physical Higgs boson, and  $v$ ($\simeq 246$ GeV) 
 is the vacuum expectation value.
The dimension-six operators 
${\mathcal O}_{t1}$ and ${\mathcal O}_{Dt}$ are 
classified as 
\begin{align}
 {\mathcal O}_{t1} =& {\mathcal O}_{t1}^{v^2\phi\bar\psi\psi}
                    +{\mathcal O}_{t1}^{v\phi^2\bar\psi\psi}
                    +{\mathcal O}_{t1}^{\phi^3\bar\psi\psi}, \label{eq:dec-Ot1}\\
 {\mathcal O}_{Dt} =& {\mathcal O}_{Dt}^{\partial\phi\bar\psi\partial\psi}
                    +{\mathcal O}_{Dt}^{vV\bar\psi\partial\psi}
                    +{\mathcal O}_{Dt}^{vV^2\bar\psi\psi}
                    +{\mathcal O}_{Dt}^{V\phi\bar\psi\partial\psi}
                    +{\mathcal O}_{Dt}^{V\partial\phi\bar\psi\psi} 
                    +{\mathcal O}_{Dt}^{V^2\phi\bar\psi\psi}, \label{eq:dec-ODt}
 \end{align}
where the concrete expressions of terms in the r.h.s. are given in Appendix.
 
\begin{figure}[t]
\includegraphics[width=8cm]{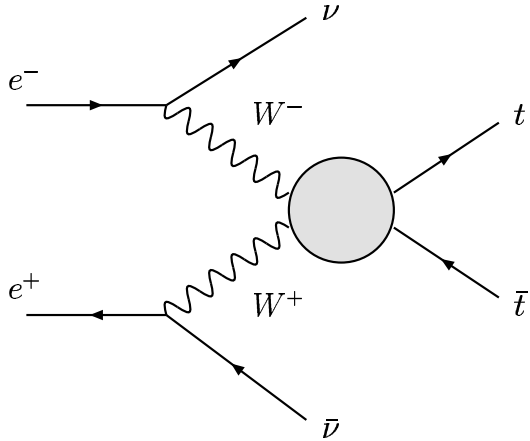}
\caption{
The top pair production via  $W$ boson fusion 
}
\label{fig:diagram-full}
\end{figure}

After the symmetry breaking, the first operator ${\mathcal O}_{t1}$
directly modifies the strength of the Yukawa coupling of 
$t \bar t H$ without changing the relation between the top quark mass 
and the SM top-Yukawa coupling.  
The second one ${\mathcal O}_{Dt}$, which includes the covariant 
derivative $D_\mu$, describes the momentum dependence of the effective
top-Yukawa coupling.
These two operators determine the behavior of 
the effective top-Yukawa coupling below $\Lambda$, 
so that it is important to separately measure 
these operators to determine the direction of  
fundamental theories which describe higher scales above $\Lambda$. 
The effect of these dimension-six operators has already been 
discussed for the process 
$e^- e^+ \to t \bar{t} H$ by Han et al\cite{han}. 
In this paper, we extend their study and examine the effect of these operators for the 
process $e^-e^+ \to W^-W^+ \nu\bar \nu \to  t \bar t \nu \bar \nu$: 
see Fig.~\ref{fig:diagram-full}.

We note that ${\mathcal O}_{t1}$ yields the 
anomalous dimension-five interaction $W^-W^+ \bar t t$ 
in the large Higgs-boson mass limit via the 
$s$-channel process  $W^-W^+ \to H^{\ast} \to t \bar t$.  
In Ref.~\cite{larios}, impact of dimension-five 
anomalous couplings 
$W^- W^+  \bar t t$ and 
$W_\mu^- W_\nu^+ \bar t \sigma^{\mu\nu}  t$ 
is studied on the cross section of $W^-W^+ \to t \bar t$ 
at an $e^- e^+$ linear collider. 
The tensor operator $W_\mu^- W_\nu^+ \bar t \sigma^{\mu\nu} t$ 
is not relevant to the new interaction between the 
Higgs boson and the top quark.    
In this paper, we concentrate on the effect of 
${\mathcal O}_{t1}$ and ${\mathcal O}_{Dt}$ with 
the Higgs boson mass $m_H^{}$ to be at the electroweak scale and the 
SM cutoff $\Lambda$ to be at TeV scales.

Let us discuss possible allowed values for the   
anomalous couplings $C_{t1}$ and $C_{Dt}$ under  
theoretical consistencies and current experimental data. 
The coefficients $C_i/\Lambda^2$ of the dimension-six 
operators ${\cal O}_i$ can be constrained theoretically by using  
the idea of partial wave unitarity\cite{pwu}. 
Due to the structure of a dimension-six operator, 
the two-body elastic scattering amplitudes is 
proportional to the square of the scattering energy, 
so that the coefficient becomes strong at some 
energy scale, and violate tree-level unitarity. 
The unitarity bounds for the coefficients $C_i/\Lambda^2$ 
are obtained by setting the scale of unitarity violation to be 
above $\Lambda$.  
The bounds for $C_{t1}$ and $C_{Dt}$ are evaluated as
\begin{align}
&\left| C_{t1} \right| \le \frac{16\pi}{3\sqrt{2}} 
\left( \frac{\Lambda}{v} \right), 
\label{eq:pu-t1} \\
&-6.2 \le C_{Dt} \le 10.2.\label{eq:pu-Dt}
\end{align}
These results are almost the same as those in Ref.~\cite{gounaris2}.

\begin{figure}[t]
\rotatebox{0}{\includegraphics[width=9cm]{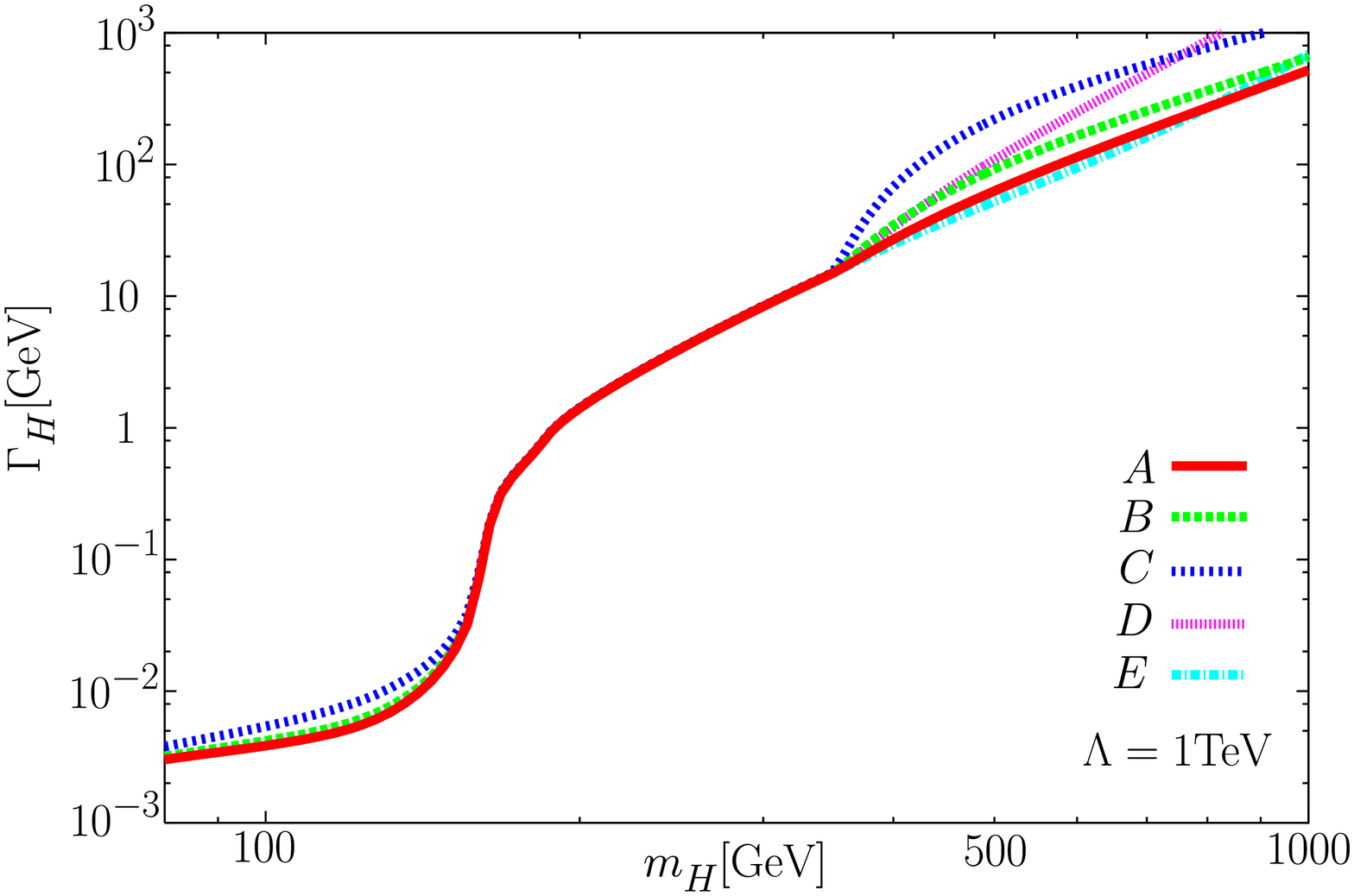}}
\caption{
The total width of the Higgs boson for several   
cases of $C_{t1}^{}$ and $C_{Dt}^{}$: $(C_{t1}^{}, C_{Dt}^{})=(0,0)$[Set
 A],
  $(+16\pi \Lambda/(3\sqrt{2}v),0)$[Set B], $(-16\pi
  \Lambda/(3\sqrt{2}v),0)$[Set C],
  $(0,+10.2)$[Set D] and $(0,-6.2)$[Set E]: see Eqs.~(\ref{eq:pu-t1}) and
  (\ref{eq:pu-Dt}). $\Lambda$ is set to be 1 TeV. 
}
\label{fig:plot-width}
\end{figure}

\begin{figure}[t]
\begin{minipage}{10cm}
\unitlength=1cm
\rotatebox{0}{
\includegraphics[width=8.5cm]{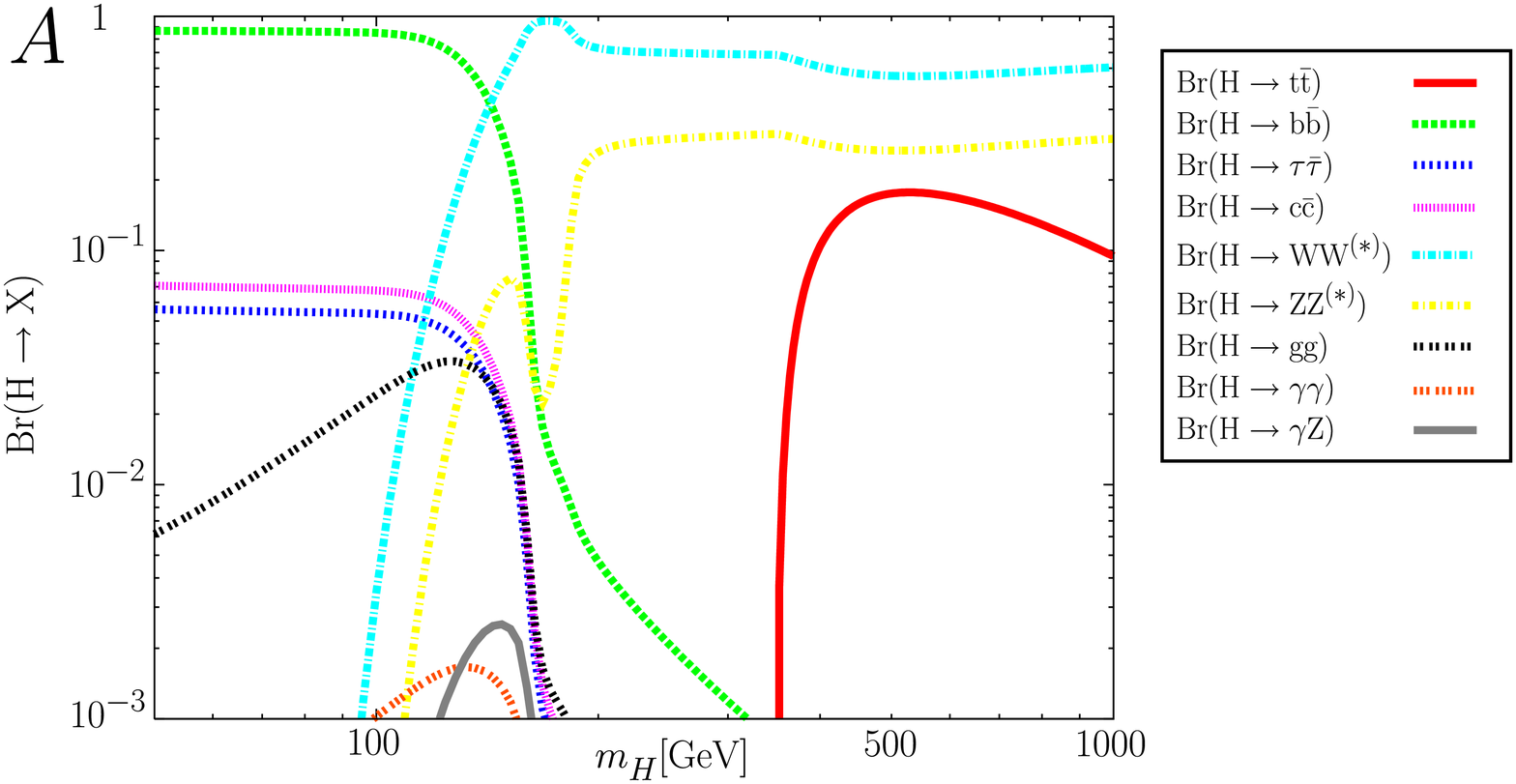}}
\end{minipage}\\
\begin{minipage}{8cm}
\unitlength=1cm
\rotatebox{0}{
\includegraphics[width=6.5cm]{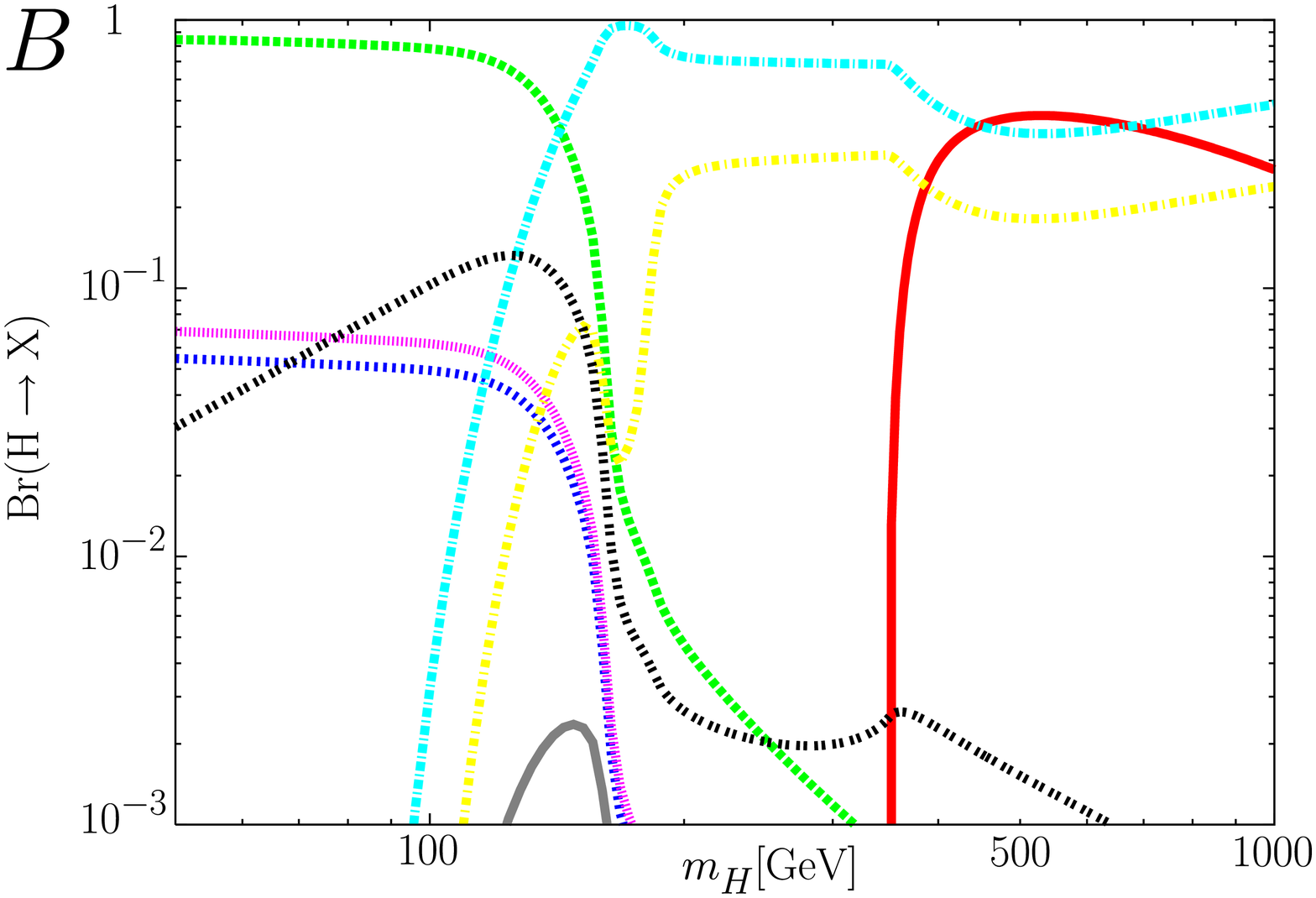}}
\end{minipage}
\begin{minipage}{8cm}
\unitlength=1cm
\rotatebox{0}{
\includegraphics[width=6.5cm]{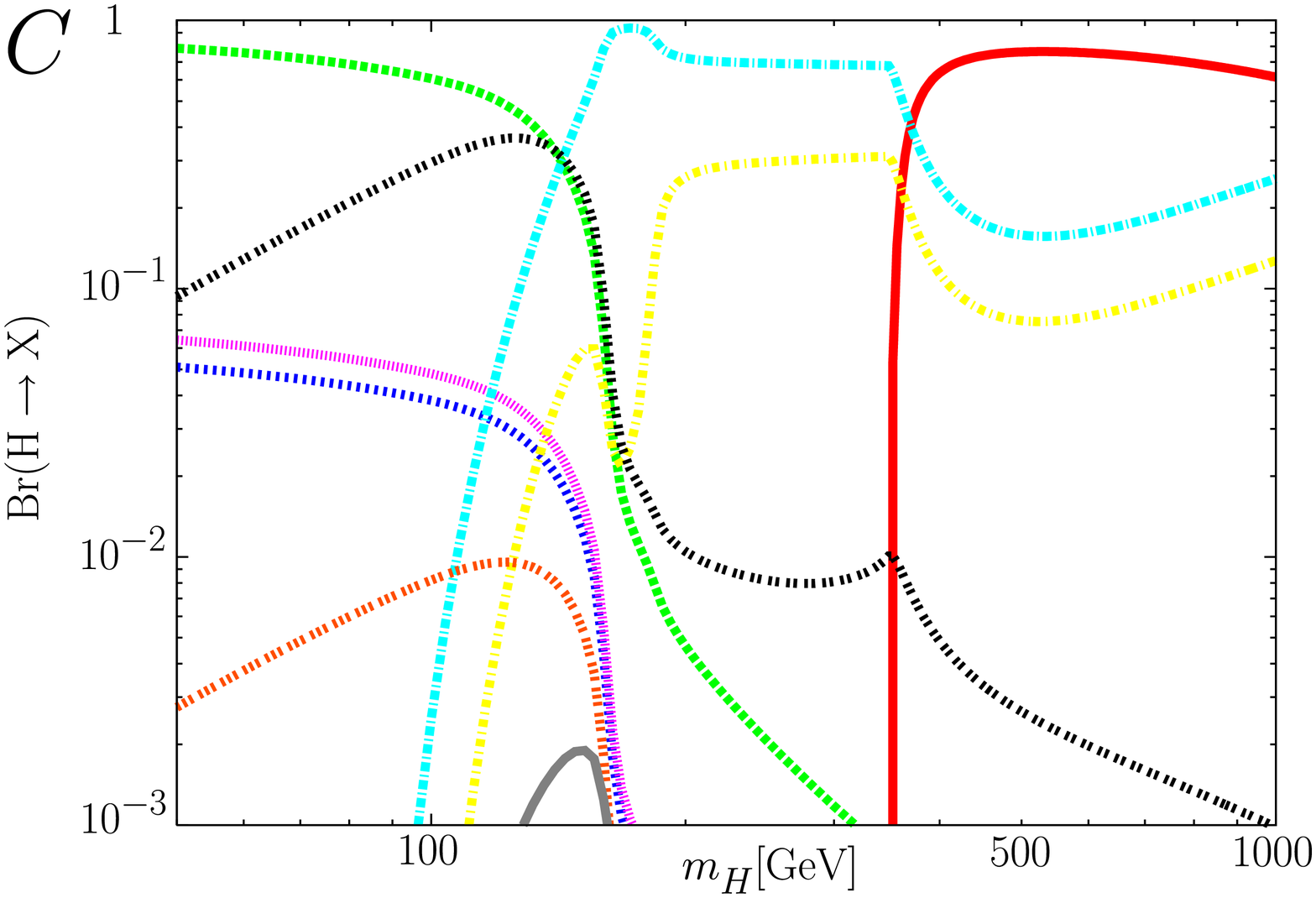}}
\end{minipage}
\begin{minipage}{8cm}
\unitlength=1cm
\rotatebox{0}{
\includegraphics[width=6.5cm]{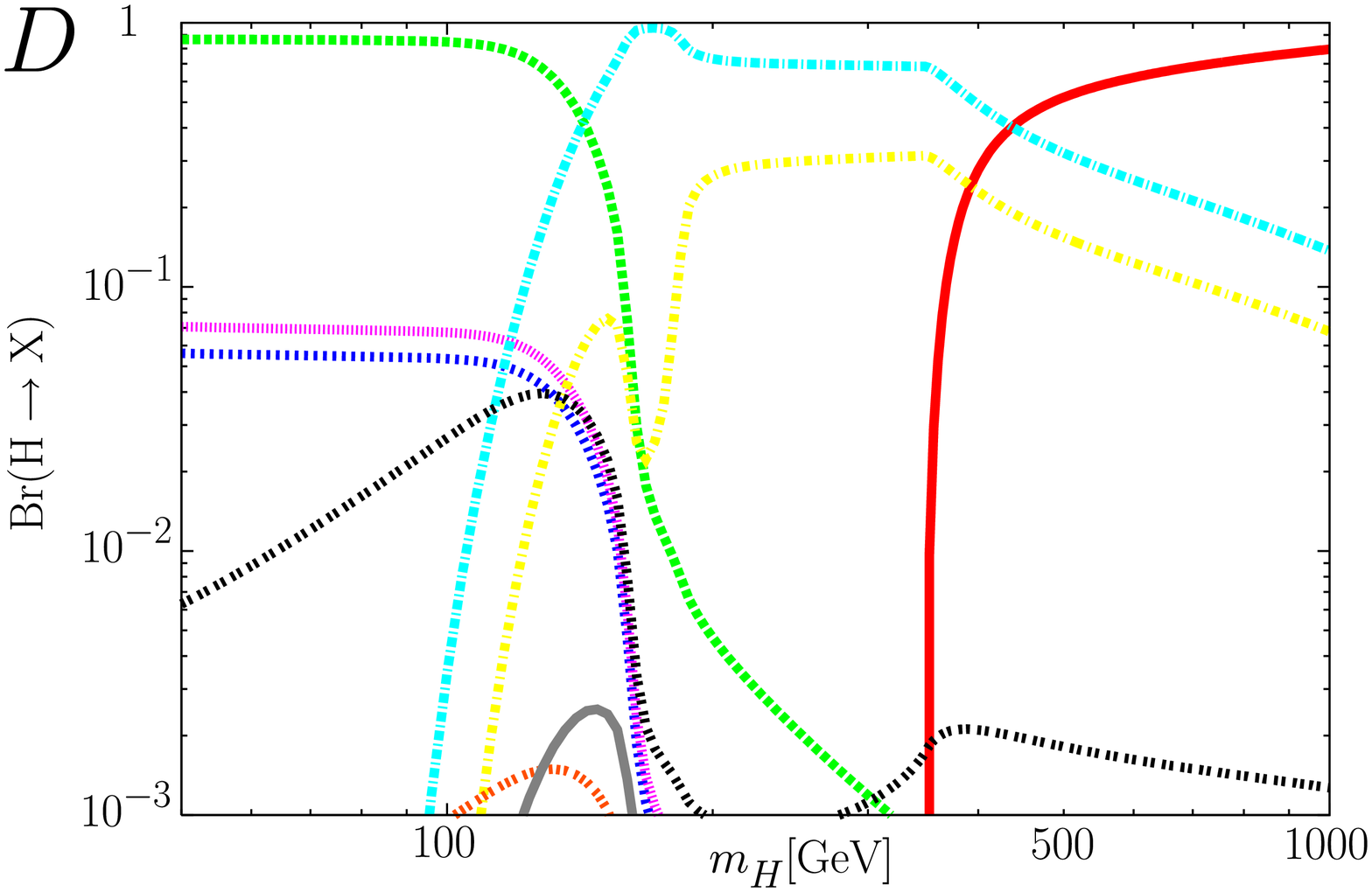}}
\end{minipage}
\begin{minipage}{8cm}
\unitlength=1cm
\rotatebox{0}{
\includegraphics[width=6.5cm]{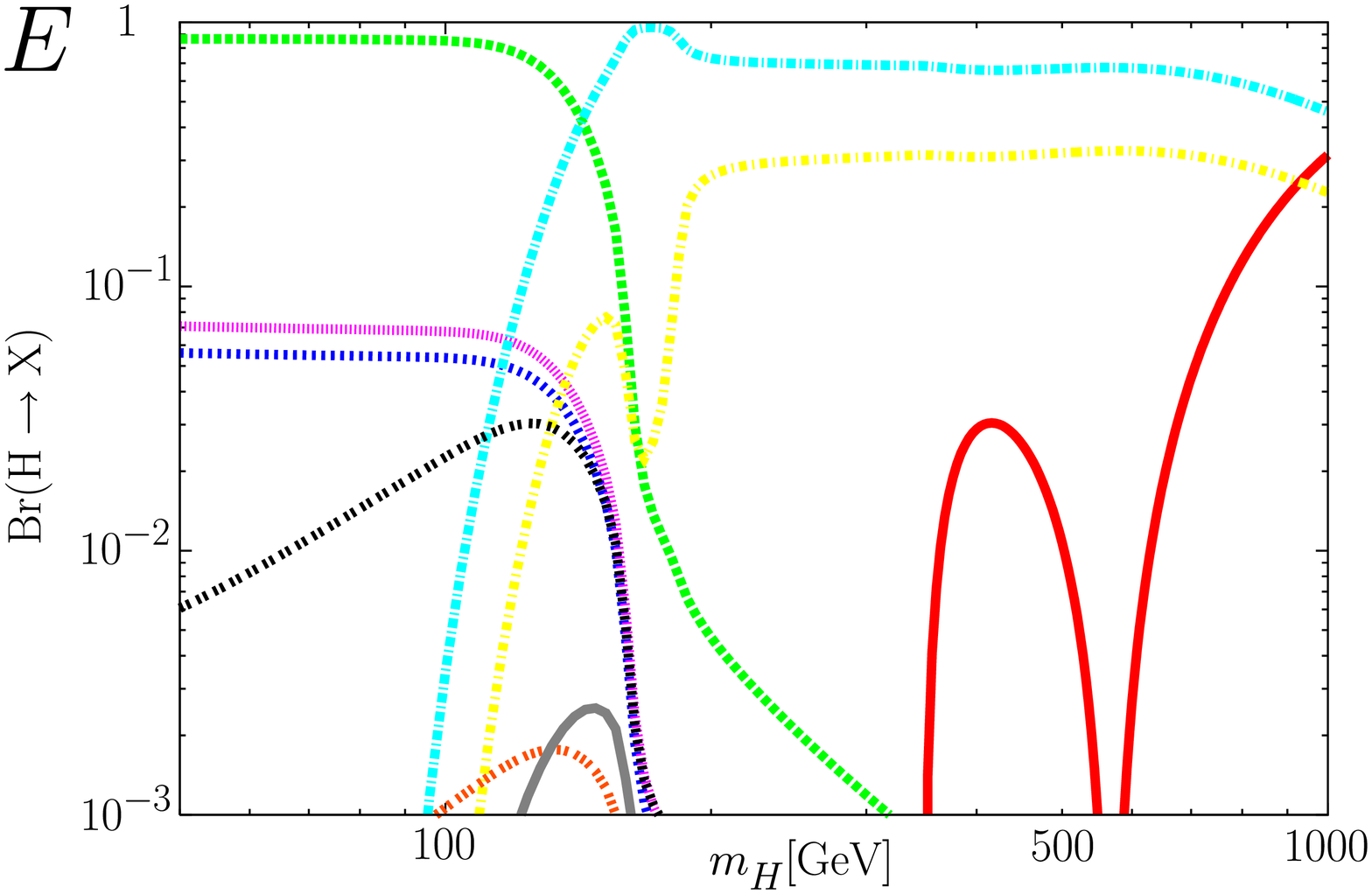}}
\end{minipage}
\caption{
Decay branching ratios of the Higgs boson for the cases of 
 $(C_{t1}^{}, C_{Dt}^{})=(0,0)$[Set
 A],
  $(+16\pi \Lambda/(3\sqrt{2}v),0)$[Set B], $(-16\pi
  \Lambda/(3\sqrt{2}v),0)$[Set C],
  $(0,+10.2)$[Set D] and $(0,-6.2)$[Set E]: see Eqs.~(\ref{eq:pu-t1}) and
  (\ref{eq:pu-Dt}). $\Lambda$ is set to be 1 TeV.
  }
\label{fig:plot-br}
\end{figure}

The value of the anomalous coupling $C_{t1}$ is 
free from the constraint from current experimental data\cite{pdg,hikasa},
because it only affects the genuine interaction between 
the top quark and the Higgs boson which has not 
been measured yet. Therefore, 
only the theoretical consideration such as 
perturbative unitarity is important to constrain this 
operator. 
On the other hand, $C_{Dt}$ turns out to  
receive strong experimental limits 
from the electroweak rho parameter result\cite{pdg,lep,Barbieri:1999tm}, since
the operator ${\cal O}_{Dt}$ changes the 
interaction of the top quark with weak gauge bosons 
through the covariant derivative.
The contribution of ${\cal O}_{Dt}^{}$ to the
rho parameter is calculated as 
\begin{align}
  \Delta \rho_{Dt}^{} &= -
 \frac{N_c}{16\pi^2}\left(\frac{m_t^2}{\Lambda^2}\right)
 \left[
 \frac{\sqrt{2} m_t}{v}C_{Dt}^{}
  \left\{
  1 - \ln \frac{\Lambda^2+m_t^2}{m_t^2} - \frac{m_t^2}{\Lambda^2 +m_t^2}
 \right\}  \right.
 \nonumber \\
& \left. +C_{Dt}^2 \left\{
 1+ \frac{m_t^2}{\Lambda^2} \left( 1 - 2 \ln
 \frac{\Lambda^2+m_t^2}{m_t^2} 
 \right)
 - \frac{m_t^2}{\Lambda^2}\frac{m_t^2}{\Lambda^2+m_t^2}
 \right\}
 \right]. \label{eq:rho-dt}
\end{align}
When $\Lambda=1$TeV, this can give a positive contribution only for 
$0 \lsim C_{Dt}^{} \lsim 3$, and its maximal value 
is $\Delta\rho_{Dt}^{} \simeq +0.001$ at $C_{Dt} \sim 1.5$.
This fact affects the allowed region of the Higgs boson mass.
For a Higgs boson mass to be 500GeV, $C_{Dt}$ is only allowed to be
around $+1.5$, where excessive negative (logarithmic) contribution of
the Higgs boson to the rho parameter is approximately
canceled by $\Delta\rho_{Dt}^{}$. 
In the following sections, we discuss the production cross section
of $e^-e^+\to \nu\bar \nu t \bar t$ and that of
its subprocess $W^+W^- \to t \bar t$.
The mass of the Higgs boson is considered typically to be
500 GeV there, so that $C_{Dt}$ is necessarily constrained
to be around +1.5 under the rho parameter result. 
Nevertheless some results will be shown for wider range of 
values of $C_{Dt}^{}$ only under the perturbative unitarity constraint.
This is for avoiding excessive exclusion of a possibility
that a combined contribution of the other dimension six
operators, which we do not discuss directly here,
compensates the positive effect of $\Delta\rho_{Dt}^{}$.
Under the rho parameter constraint, the effect of $C_{Dt}^{}$
becomes much smaller than that of $C_{t1}$.  

  The width of the Higgs boson $\Gamma_H^{}$ is evaluated
  by calculating the decay rates for $H\to f\bar f$ ($f$: quarks and
  charged leptons), $WW^{(\ast)}$,
  $ZZ^{(\ast)}$, $\gamma\gamma$, $Z\gamma$ and $gg$.
  At the leading order, they are given from the SM result
  by replacing the top Yukawa coupling by
  $y^{\rm eff}_t(m_H^2,\Lambda)$, where the effective coupling
  $y^{\rm eff}_t(-q^2,\Lambda)$ is defined by 
  \begin{align}
y^{\rm eff}_t(-q^2,\Lambda) =
 y_t^{\rm SM}
 - v^2 \frac{C_{t1}}{\Lambda^2} 
- q^2 \frac{C_{Dt}}{2 \Lambda^2}, \label{eq:yeff}
\end{align}
where $y_t^{\rm SM}$ ($= \sqrt{2} m_t/v$) is the top Yukawa coupling 
of the SM, and the rest is the additional contributions 
of the dimension-six operators. 
  In Fig.~\ref{fig:plot-width}, the values of the
  total width of the Higgs boson are plotted for each set of
  $C_{t1}^{}$ and $C_{Dt}^{}$: $(C_{t1}^{}, C_{Dt}^{})=(0,0)$ [Set A],
  $(+16\pi \Lambda/(3\sqrt{2}v),0)$ [Set B], $(-16\pi
  \Lambda/(3\sqrt{2}v),0)$ [Set C],
  $(0,+10.2)$ [Set D] and $(0,-6.2)$ [Set E], according to the
  unitarity bounds in Eqs.~(\ref{eq:pu-t1}) and
  (\ref{eq:pu-Dt}).
  Set A corresponds to the SM case. 
  The decay modes $H\to t\bar t$ (tree), $H \to
  \gamma\gamma$, $H \to \gamma Z$ and  $H \to gg$ (one loop)
  are largely modified at the leading order by the inclusion of
  ${\cal O}_{t1}$ and ${\cal O}_{Dt}$. 
  The results for the decay branching ratios
  are shown in Fig.~\ref{fig:plot-br} for Set A, Set B,
  $\cdot\cdot\cdot$, and Set E.
    
\begin{figure}[t]
\includegraphics[width=14cm]{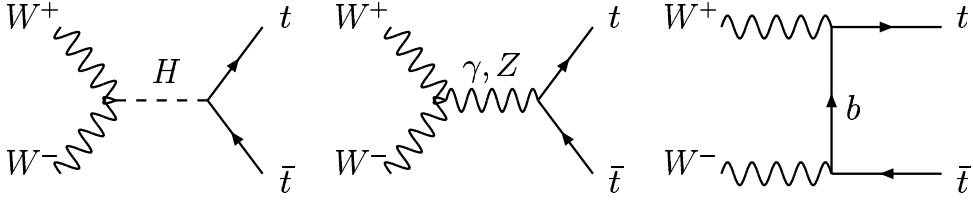}
\caption{
 Feynman diagrams 
 for the subprocess $W^-W^+ \to t \bar t$
 in the SM.
}
\label{fig:diagram-sm}
\end{figure}

\begin{figure}[t]
\rotatebox{0}{\includegraphics[width=10cm]{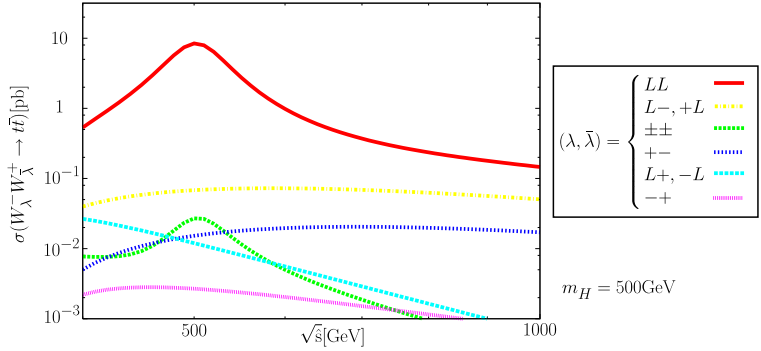}}
\caption{
 The helicity cross sections for $W^-_\lambda W^+_{\bar \lambda} \to t \bar t$ as a function of the collision energy $\sqrt{\hat{s}}$ in the SM,
 where $\lambda$ ($\bar \lambda$) is the helicity of the incoming
 $W^-$ ($W^+$) boson. The mass $m_H$ of the Higgs boson is set to be $500$ GeV.
}
\label{fig:helicity-sm}
\end{figure}

\section{Subprocess}
\label{Sec:Subprocess}

We here study the cross section for the subprocess 
$W^-W^+ \to t \bar t$\cite{RuizMorales:1999kz}.
By using the results in this section,
we evaluate the cross section of the full process 
$e^-e^+ \to W^-W^+ \nu \bar \nu \to t \bar t \nu \bar \nu$
in the the effective $W$ approximation (EWA)\cite{ewa} in
Sec.~IV,  
and compare it to the numerical results of calculation of 
the full matrix elements by the full use of the packages
CalcHEP\cite{calchep} and LanHEP\cite{lanhep}.

In the SM, Feynman diagrams for the subprocess 
are shown in Fig.~\ref{fig:diagram-sm}. 
Cross sections $\hat{\sigma}_{\lambda\bar\lambda}^{\rm SM}$ for the helicity amplitudes of  
$W_\lambda^- W_{\bar \lambda}^+ \to t\bar{t}$ 
with helicity sets ($\lambda$, $\bar \lambda$) 
are shown as a function of $\sqrt{\hat{s}}$ 
(the energy of the subprocess) in Fig.~\ref{fig:helicity-sm}, 
where $\lambda$ ($\bar \lambda$) is the helicity of 
the incoming $W^-$ ($W^+$) boson and  
the mass of the Higgs boson is set to be 
$500$ GeV. 
The polarization vector for the $W$ boson 
with the momentum $p^\mu=(E_W^{},0,0,p_W^{})$
($E_W^2=m_W^2+p_W^2$) 
and the helicity $\lambda$ is defined by 
\begin{eqnarray}
\epsilon_W^{\mu}(p,\lambda) = \left\{
\begin{array}{cc}
 \frac{1}{m_W^{}}(p_W^{},0,0,E_W^{})& ({\rm for\;\;} \lambda=L) \\
                 (0,1, \pm i,0) & ({\rm for\;\;} \lambda=\pm) \\
\end{array} \right. .
\end{eqnarray}

\noindent
It can be seen that the contribution from the 
longitudinally polarized $W$ bosons, $\hat{\sigma}_{LL}^{\rm SM}$ 
dominates over those from the other polarization sets 
in a wide region of $\sqrt{\hat{s}}$. 
In particular, around the scale of the Higgs boson mass, 
$\hat{\sigma}_{LL}^{\rm SM}$ is about one or two 
orders of magnitude greater than $\hat{\sigma}_{L-}^{\rm SM}$, the second 
largest mode, because of the $s$-channel Higgs-boson resonance. 

Next, we consider the the cross section 
in the model with the dimension-six operators
${\cal O}_{t1}^{}$ and ${\cal O}_{Dt}^{}$.
The Feynman diagrams are the same as those in the SM in
Fig.~\ref{fig:diagram-sm}, but
${\cal O}_{t1}$ and ${\cal O}_{Dt}$ 
change the coupling strength of $t\bar t H$, $Z t \bar t$ and $W^\pm t b$  
from their SM values. The diagrams for the Higgs boson 
mediation, the $Z$ boson mediation and the $b$-quark mediation 
are respectively given by 
\begin{align}
&{\mathcal M}_H^{WW\to t\bar{t}} \times
 (2\pi)^3\sqrt{2p^0}\sqrt{2{p'}^0} 
=\frac{(2 m_W^2/v)}
{\hat{s}-m_H^2+i m_H^{} \Gamma_H^{}}
 \frac{y_t^{\rm eff}(\hat{s},\Lambda)}{\sqrt{2}}
(\epsilon_\lambda \cdot {\epsilon'}_{\bar \lambda}) \bar{u}_t v_t, \label{h-med}\\
&{\mathcal M}_Z^{WW\to t\bar{t}} 
\times (2\pi)^3\sqrt{2p^0}\sqrt{2{p'}^0}=
- \frac{2 m_W^2/v^2}{\hat{s}-m_Z^2} 
A^\mu_{\lambda \bar \lambda} \bar{u}_t \left[ i \gamma_\mu 
(V_t+A_t \gamma_5) + \frac{C_{Dt}}{\Lambda^2} 
\frac{v}{2\sqrt{2}} K_\mu \right]v_t,  \\
&{\mathcal M}_b^{WW\to t\bar{t}} 
\times (2\pi)^3\sqrt{2p^0}\sqrt{2{p'}^0}&  \nonumber \\
 & \hspace*{+1cm}
=-\frac{2m_W^2/v^2}{\hat{u}-m_b^2} 
\epsilon_\lambda^\mu {\epsilon'}_{\bar \lambda}^\nu \bar{u}_t \left[
\left(i \gamma_\nu + \frac{C_{Dt}}{\sqrt{2}} \frac{v}{\Lambda^2}
k_\nu \right)
 P_L i \gamma_\rho (p-k')^\rho
 \left( i \gamma_\mu - \frac{C_{Dt}}{\sqrt{2}}
 \frac{v}{\Lambda^2}
 {k}'_\mu \right) \right] v_t,  
\end{align}
where $\epsilon=\epsilon(p,\lambda)$ and
$\epsilon'=\epsilon(p',\bar\lambda)$; $\bar u_t$ ($v_t$) is the
Dirac spinor for outgoing $t$ ($\bar t$); 
$p$ ($p'$) and $k$ ($k'$) are the momenta
of the incoming $W^-$ ($W^+$) and outgoing
$t$ ($\bar t$); $q=p+p'$, $P=p-p'$, $K=k-k'$,
$P_{L}^{}=(1 + \gamma_5)/2$,
 $A_t=\frac{1}{2}$, $V_t=\frac{1}{2}-2 Q_t \sin^2\theta_W$ with $\theta_W^{}$
 being the Weinberg angle; and
  $A_{\lambda\bar\lambda}^\mu = (\epsilon_\lambda \cdot
  {\epsilon'}_{\bar \lambda})P^\mu
  + 2 (\epsilon_\lambda \cdot q){\epsilon'}_{\bar \lambda}^\mu
  - 2({\epsilon'}_{\bar \lambda} \cdot q) \epsilon_\lambda^\mu $.
  Mandelstam variables\footnote{We use the metric
  $g_{\mu\nu}={\rm diag}(-1,+1,+1,+1)$, and our
  definition of $C_{Dt}$ is different from
  that in Ref.~\cite{gounaris1} by a factor of $-1$.
  } are defined by
  $\hat{s}=-q^2$, $\hat{t}=-(p-k)^2=-(p'-k')^2$ and
  $\hat{u}=-(p-k')^2=-(p'-k)^2$.
The diagram of the photon mediation is not changed by the 
existence of the operators ${\cal O}_{t1}$ and ${\cal O}_{Dt}^{}\,$; 
\begin{align}
{\mathcal M}_\gamma^{WW\to t\bar{t}} 
\times (2\pi)^3\sqrt{2p^0}\sqrt{2{p'}^0}=& 
- \frac{Q_t e^2}{\hat{s}} A^\mu_{\lambda \bar\lambda} 
\bar{u}_t i \gamma_\mu v_t.
\end{align}

\begin{figure}[t]
\includegraphics[width=14cm]{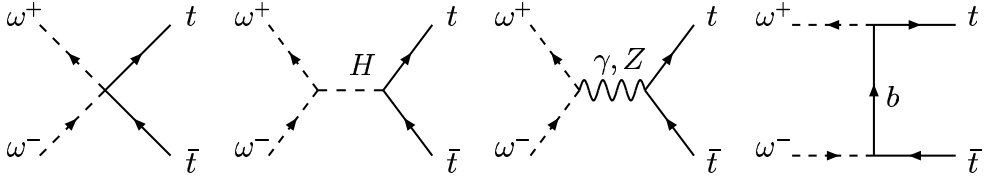}
\caption{
  Feynman diagrams for the subprocess
 $\omega^-\omega^+ \to t \bar t$ 
 in the model with the dimension-six operators
 ${\cal O}_{t1}$ and ${\cal O}_{Dt}$.
}
\label{fig:wwtt-sm}
\end{figure}

As seen from Eq.~(\ref{h-med}), the operator ${\cal O}_{t1}$ 
represents the shift of the SM top-Yukawa coupling, while 
${\cal O}_{Dt}$ gives its momentum dependence.  
The effect of ${\cal O}_{Dt}$ enhances the effective 
Yukawa coupling as $\sqrt{\hat{s}}$ grows until the cutoff 
scale $\Lambda$. It turns out that 
this momentum dependence of the effective Yukawa coupling
due to ${\cal O}_{Dt}$ in the Higgs boson mediation 
is replaced by the Higgs boson mass dependence
due to the gauge cancellation mechanism. 
The invariant amplitude of the subprocess is obtained as 
\begin{align}
    {\cal M}^{WW\to tt} = 
{\cal M}^{WW\to tt}_H
+{\cal M}^{WW\to tt}_b
+{\cal M}^{WW\to tt}_Z
+{\cal M}^{WW\to tt}_\gamma. 
\end{align}

Calculation of the amplitude can be tested 
by using the equivalence of 
the longitudinally polarized weak boson ($W^\pm_L$) and the corresponding 
Nambu-Goldstone boson ($\omega^\pm$) for $\sqrt{\hat{s}} \gg m_W^{}$\cite{et}. 
In the Feynman gauge,  the diagrams for
the process $\omega^- \omega^+ \to t \bar t$ are shown
in Fig.~\ref{fig:wwtt-sm}, and each diagram is calculated as 
\begin{align}
{\mathcal M}_\times^{\omega^- \omega^+ \to t \bar t} 
(2 \pi)^3 \sqrt{2p^0} \sqrt{2{p'}^0} & = -\frac{C_{t1}}{\Lambda^2}
 \frac{v}{\sqrt2} \bar u_t v_t, \\
{\mathcal M}_H^{\omega^- \omega^+ \to t \bar t} 
(2 \pi)^3 \sqrt{2p^0} \sqrt{2{p'}^0} & = \frac{m_H^2/v}{\hat{s}-m_H^2+i
 m_H \Gamma_H}
\frac{y_t^{\rm eff}(\hat{s},\Lambda)}{\sqrt{2}}
 \bar u_t   v_t, \\
{\mathcal M}_\gamma^{\omega^- \omega^+ \to t \bar t} 
(2 \pi)^3 \sqrt{2p^0} \sqrt{2{p'}^0} & = - \frac{Q_t e^2}{\hat{s}}  \bar
 u_t i P\!\!\!\!/\, v_t, \\
{\mathcal M}_Z^{\omega^- \omega^+ \to t \bar t} 
(2 \pi)^3 \sqrt{2p^0} \sqrt{2{p'}^0} &
 = - \frac{g_Z^2 (V_t + A_t)}{4 (\hat{s}-m_Z^2)} \bar u_t  \left( i P\!\!\!\!/\,
 (V_t + A_t \gamma_5)  + \frac{C_{Dt}}{\Lambda^2} \frac{v}{2\sqrt2}
 K\cdot P \right) v_t, \\
{\mathcal M}_b^{\omega^- \omega^+ \to t \bar t} 
(2 \pi)^3 \sqrt{2p^0} \sqrt{2{p'}^0} & = - \frac{1}{\hat{u}-m_b^2}
 \left(y_t^{\rm SM} - \frac{C_{Dt}}{\Lambda^2} p\cdot {k'}
 \right)^2
 \bar u_t P_L  i (p\!\!\!\!/\,-k'\!\!\!\!/\,) v_t, 
\end{align}
where momenta $p$ ($p'$) and $k$ ($k'$) are 
defined in a similar way to the case of $W^-W^+\to t \bar t$.
The invariant amplitude is obtained as 
\begin{align}
    {\cal M}^{\omega\omega\to tt} = 
{\cal M}^{\omega\omega\to tt}_{\times} 
+{\cal M}^{\omega\omega\to tt}_H
+{\cal M}^{\omega\omega\to tt}_\gamma
+{\cal M}^{\omega\omega\to tt}_Z
+{\cal M}^{\omega\omega\to tt}_b. 
\end{align}

\begin{figure}[t]
\begin{minipage}{8cm}
\unitlength=1cm
\rotatebox{0}{
\includegraphics[width=7.5cm]{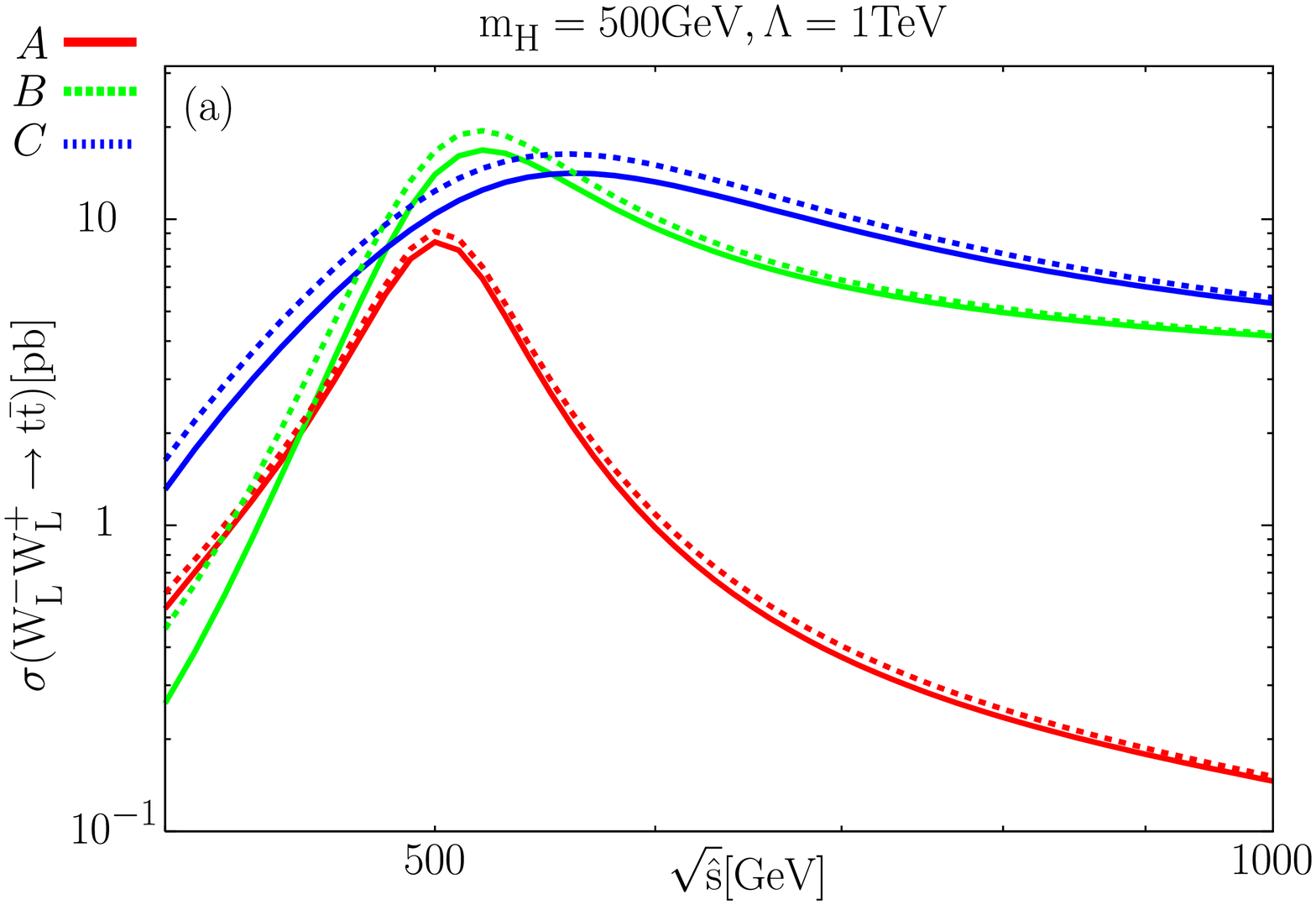}}
\end{minipage}
\begin{minipage}{8cm}
\unitlength=1cm
\rotatebox{0}{
\includegraphics[width=7.5cm]{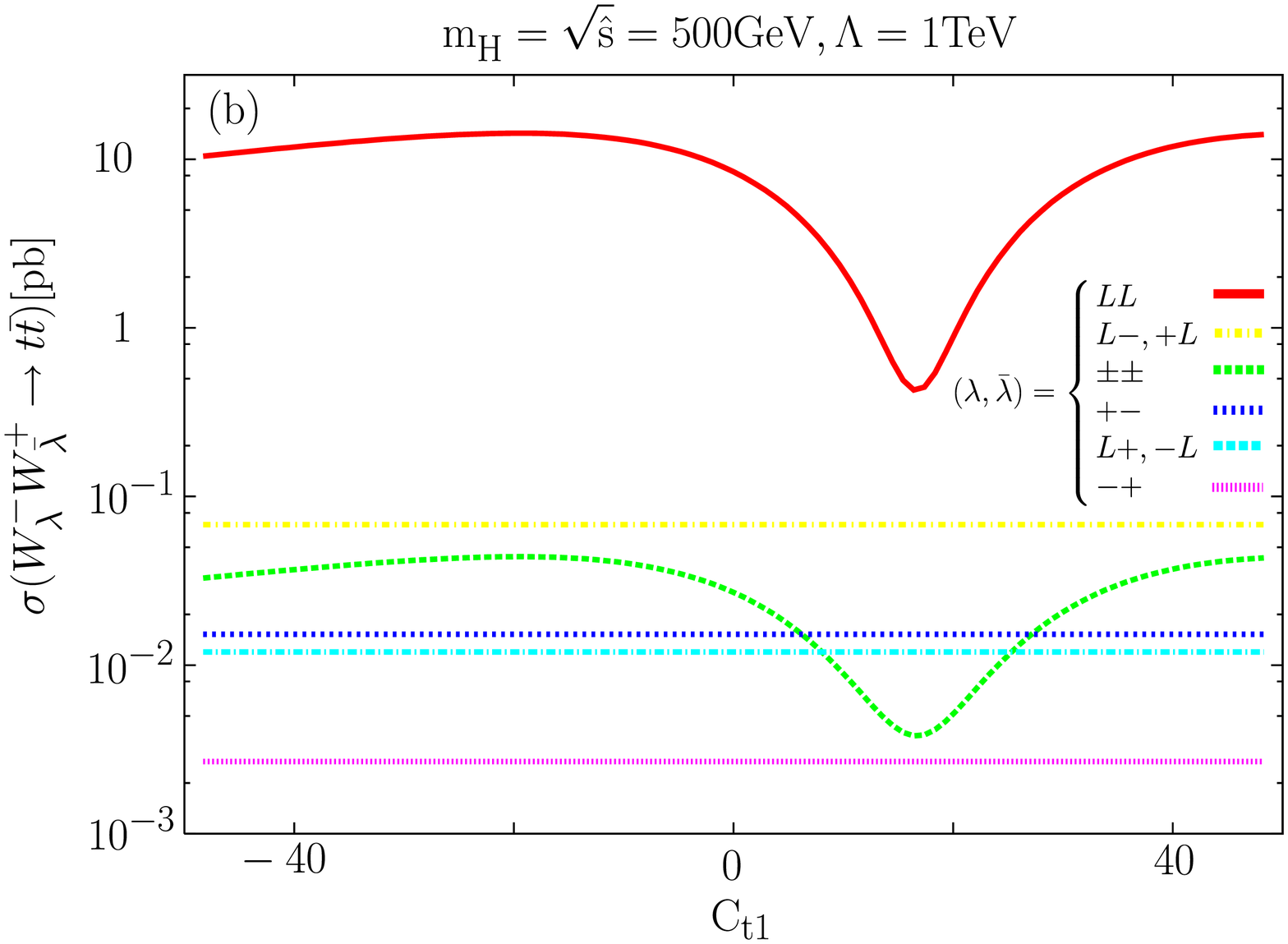}}
\end{minipage}
\caption{(a) Cross sections of $W^-_L W^+_L \to t \bar t$ (solid curves)
  and $\omega^-\omega^+\to t\bar t$ (dotted curves) as a function
  of $\sqrt{\hat{s}}$ for $m_H^{}=500$GeV and $\Lambda=1$TeV.
  The magnitudes of $C_{t1}$ are set corresponding to the upper and
  lower limit from perturbative unitarity (Set B and Set C). The SM
  results are also shown.
  (b) Cross sections of $W^-_{\lambda} W^+_{\bar\lambda} \to t \bar t$
  as a function of $C_{t1}$ for $\sqrt{\hat{s}}=m_H^{}=500$GeV and $\Lambda=1$TeV. }
\label{fig:WW-tt-ct1}
\end{figure}

\begin{figure}[t]
\begin{minipage}{8cm}
\unitlength=1cm
\rotatebox{0}{
\includegraphics[width=7.5cm]{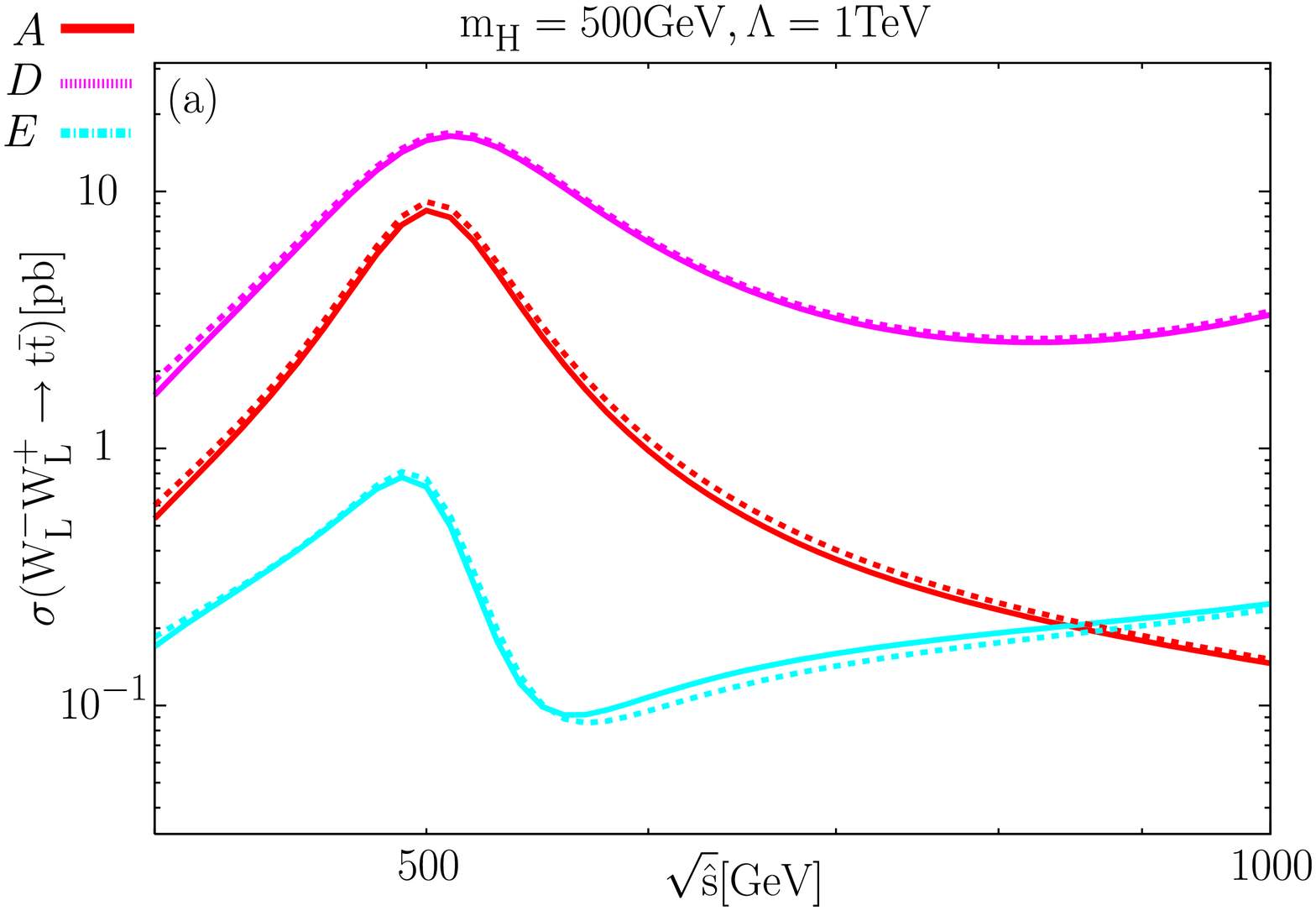}}
\end{minipage}
\begin{minipage}{8cm}
\unitlength=1cm
\rotatebox{0}{
\includegraphics[width=7.5cm]{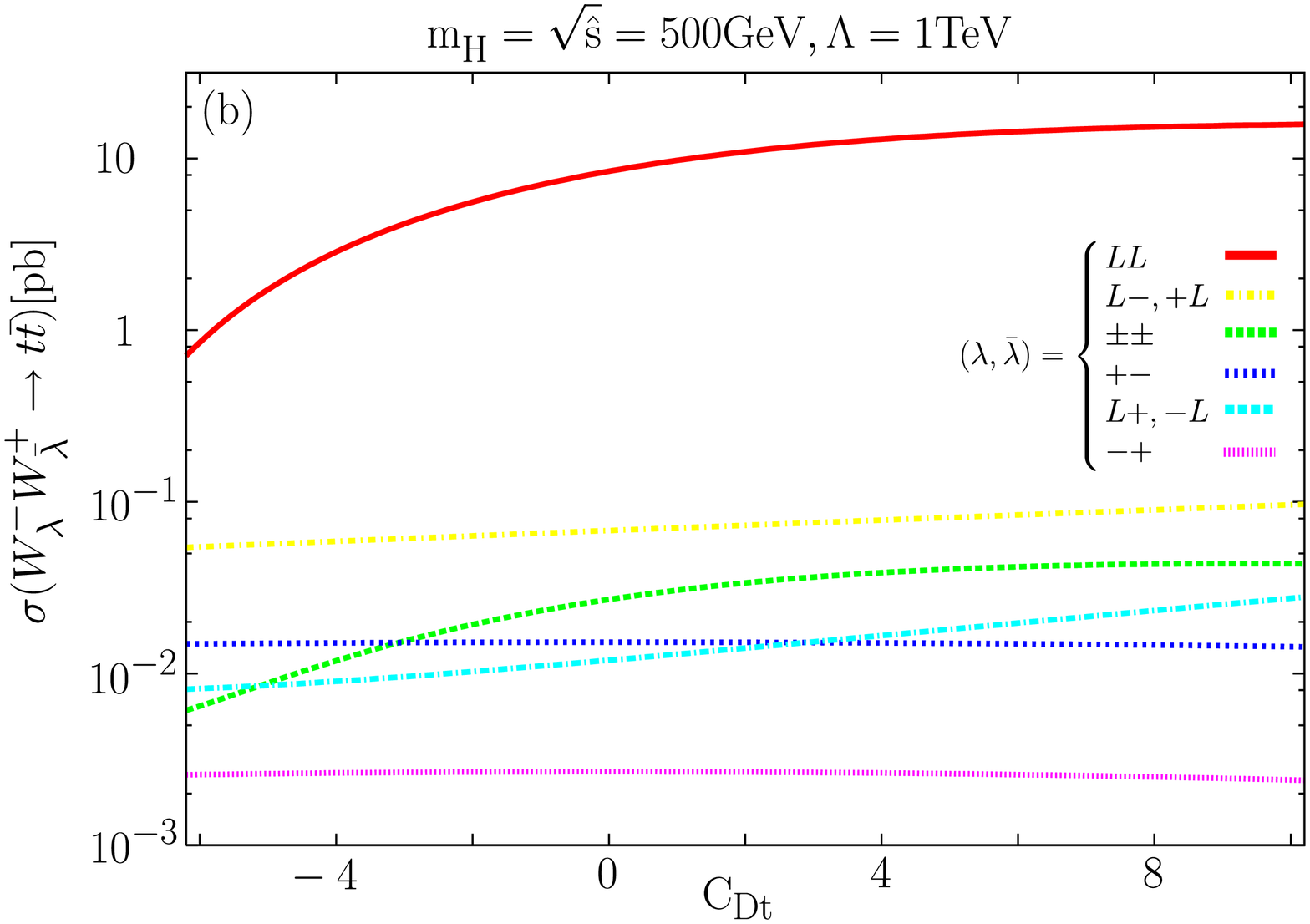}}
\end{minipage}
\caption{(a) Cross sections of $W^-_L W^+_L \to t \bar t$ (solid curves)
  and $\omega^-\omega^+\to t\bar t$ (dotted curves) as a function
  of $\sqrt{\hat{s}}$ for $m_H^{}=500$GeV and $\Lambda=1$TeV.
  The magnitudes of $C_{Dt}$ are set corresponding to the upper and
  lower limit from perturbative unitarity (Set D and Set E). The SM
  results (Set A) are also shown.
  (b) Cross sections of $W^-_{\lambda} W^+_{\bar\lambda} \to t \bar t$
  as a function of $C_{Dt}$ for $\sqrt{\hat{s}}=m_H^{}=500$GeV and
  $\Lambda=1$TeV.
}
\label{fig:WW-tt-cDt}
\end{figure}

The numerical results for the subprocess 
$W_\lambda^-W_{\bar \lambda}^+\to t \bar t$
in the model with the dimension-six operators 
${\cal O}_{t1}$ and ${\cal O}_{Dt}$ are shown 
as solid curves
in Figs.~\ref{fig:WW-tt-ct1}(a) and  \ref{fig:WW-tt-ct1}(b) and in   
Figs.~\ref{fig:WW-tt-cDt}(a) and  \ref{fig:WW-tt-cDt}(b), respectively.
The cross sections of the corresponding
processes $\omega^-\omega^+ \to t \bar t$ are also plotted by dotted
curves for a test of the calculation by using the equivalence theorem.
It can be seen that the cross sections for
$W^-_L W^+_L \to t \bar t$ and $\omega^-\omega^+ \to t \bar t$
agree with each other at high energies $\sqrt{\hat{s}} \gg m_W^{}$.
The Higgs boson mass is fixed as $m_H=500$ GeV.


In Fig.~\ref{fig:WW-tt-ct1}(a), the coefficient of ${\cal O}_{t1}$ is taken 
to be $C_{t1} = \pm \frac{16\pi}{3\sqrt{2}} 
\left( \frac{\Lambda}{v} \right)$ (Set B and Set C), corresponding to 
the upper and lower bounds determined from perturbative unitarity 
in Eq.~(\ref{eq:pu-t1}).
The effect of the other operator ${\cal O}_{Dt}$ is switched off. 
The SM prediction (Set A) is also shown for comparison.
The cutoff scale $\Lambda$ is set to be $1$ TeV. 
It is found that the effect of ${\cal O}_{t1}$ can 
enhance the cross section from its SM prediction significantly. 
In particular, for $\sqrt{\hat{s}}$ higher than 400 GeV, 
values of the cross section can even be 
one order of magnitude greater than its SM results. 

In Fig.~\ref{fig:WW-tt-ct1}(b), the helicity cross sections for
$W^-_\lambda W^+_{\bar\lambda} \to t \bar t$
are shown as a function of the anomalous coupling $C_{t1}$
for $\sqrt{\hat s} = m_H^{}=500$ GeV with $C_{Dt}$ being switched off.
The SM prediction corresponds to the point of $C_{t1}=0$.
The dimension-six operator 
${\mathcal O}_{t1}$ only contributes to $\lambda \bar \lambda = LL,
$ and $\pm\pm$.

In Fig.~\ref{fig:WW-tt-cDt}(a),
the coefficient of ${\cal O}_{Dt}$ is taken to be $C_{Dt} = 10.2$ (Set
D) and
$-6.2$ (Set E) which are the lower and upper bounds from perturbative unitarity
in Eq.~(\ref{eq:pu-Dt}).
$C_{t1}$ is set to be zero.
The SM prediction (Set A) is also shown for comparison.
For higher $\sqrt{\hat s}$, cross sections with large dimension-six
coupling $C_{Dt}$ are enhanced due to its momentum dependence. 

In Fig.~\ref{fig:WW-tt-cDt}(b), the helicity cross sections for
$W^-_\lambda W^+_{\bar \lambda} \to t \bar t$ are plotted as a function of $C_{Dt}$
for all $W$ polarizations with $\sqrt{\hat s} = m_H^{}=500$ GeV.
Again $C_{t1}$ is switched off.

As already discussed, although $C_{Dt}$ receives relatively strong
constraint under the LEP precision data, 
$C_{t1}$ is free from them.
Therefore, we conclude that the effect of ${\cal O}_{t1}$
can be really large as compared to the SM prediction.
One might suspect why the correction due to ${\cal O}_{t1}$ 
can be so large as compared to the SM result.  
The reason is simple. 
Let us look at the effective Yukawa coupling $y_t^{\rm eff}(-q^2,\Lambda)$
given in Eq.~(\ref{eq:yeff}).
While the SM Yukawa interaction is determined 
by $y_t^{\rm SM} = \sqrt{2} m_t/v \sim {\cal O}(1)$, 
the magnitude of $C_{t1}$ in Eq.~(\ref{eq:yeff})
is only constrained by the 
unitarity bound in Eq.~(\ref{eq:pu-t1}), and its maximum contribution 
to $y_t^{\rm eff}(\hat{s},\Lambda)$ can be about $\pm 2.9$ for $\Lambda=1$ TeV. 
Therefore, $y_t^{\rm eff}(\hat{s}, \Lambda)$ can be enhanced by 
$C_{t1}$ to be about 2-4 times greater than 
$y_t^{\rm SM}$.
Consequently, the cross section 
can be about several times 100\% enhanced.
Such a large enhancement can also occur in the non-standard contribution
to the effective self-coupling of the SM-like Higgs
boson\cite{Barger:2003rs,hhh-thdm}.
On the other hand, the contribution of $C_{Dt}$ to 
$y_t^{\rm eff}(-q^2,\Lambda)$ is strongly limited if the LEP result
is taken into account, so that its effect can change
the SM Yukawa coupling by at most 10-20 \%. 
 
\section{Cross section of the full process}
\label{Sec:Cross section of the full process}

We here evaluate the cross section of the full process 
$e^-e^+ \to W^-W^+ \nu \bar \nu \to t \bar t \nu \bar \nu$ 
in the method of the EWA\cite{ewa} 
by using the result of calculation of the subprocess.
We also show the results of full matrix-element calculation
based on CalcHEP\cite{calchep} and LanHEP\cite{lanhep}, and compare the both results.
As we show soon later, the EWA gives reasonable results
for a large value of $\sqrt{\hat{s}}$ as compared to $m_W^{}$.  
In order to keep the validity of the calculation based on the EWA,
we need to make the kinematic cut at an appropriate value. 
Here we employ the cut $M_{tt}^{}>400$ GeV\cite{godfrey}.
The accuracy of the EWA has been discussed by many authors\cite{Johnson:1987tj,nagashima}.
Our results agree with those in  Ref.~\cite{larios} where
expected error is evaluated to be
of the order of 10\% for the cut $M_{tt}^{}>500$ GeV.

\begin{figure}[t]
\rotatebox{0}{
\includegraphics[width=9cm]{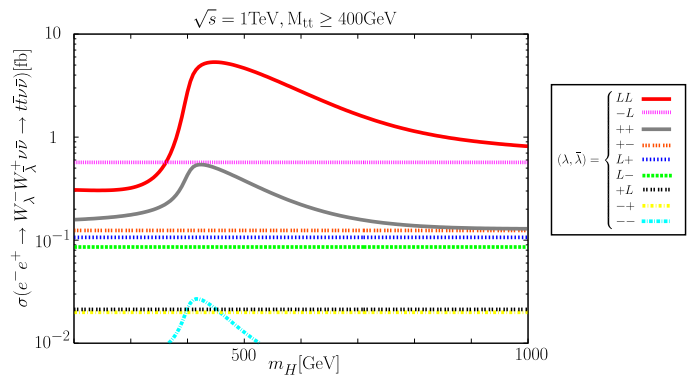}}
\caption{The cross section of
 $e^-e^+\to W^-_\lambda W^+_{\bar \lambda} \nu \bar \nu \to t \bar t \nu
 \bar \nu$ evaluated in the EWA, where $\lambda$ ($\bar \lambda$) is
 the helicity of incoming $W^-$ ($W^+$) boson. 
 } 
\label{fig:SigmaCut400-SM}
\end{figure}

 In Fig.~\ref{fig:SigmaCut400-SM}, the SM results for the cross section
 of
 $e^-e^+\to W^-_\lambda W^+_{\bar \lambda} \nu \bar \nu \to t \bar t \nu
  \bar \nu$
 are shown as a function of $m_H^{}$  
 with the kinematic cut $M_{tt} \ge 400$ GeV.
 The electron-positron collider energy $\sqrt{s}$ is set to be 1 TeV.
 This is a reproduction of the results
 in Refs.~\cite{godfrey,godfrey2}
 For a heavy Higgs boson ($m_H^{} > 2 m_t$) the longitudinally
 polarized mode ($LL$) is dominant due to the resonance of
 the Higgs boson in the Higgs boson
 mediated ($s$-channel) diagram, while for a
 relatively light Higgs boson all the modes $LL$, $LT$ and $TT$
 are comparable.
 Therefore, information of the Yukawa interaction
 can more easily be extracted for heavier Higgs bosons.

\begin{figure}[t]
\begin{minipage}{8cm}
\unitlength=1cm
\rotatebox{0}{
\includegraphics[width=7.5cm]{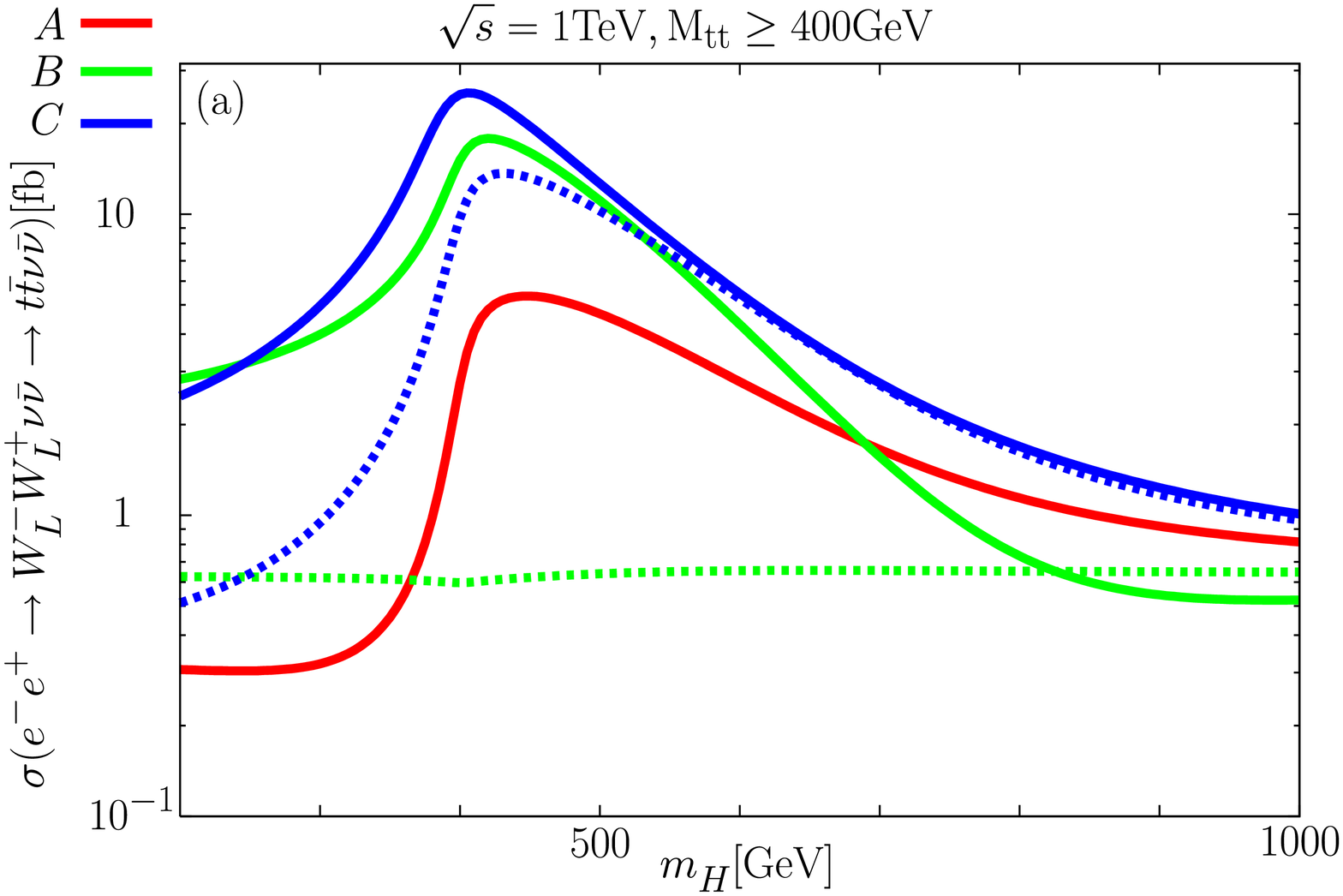}}
\end{minipage}
\begin{minipage}{8cm}
\unitlength=1cm
\rotatebox{0}{
\includegraphics[width=7.5cm]{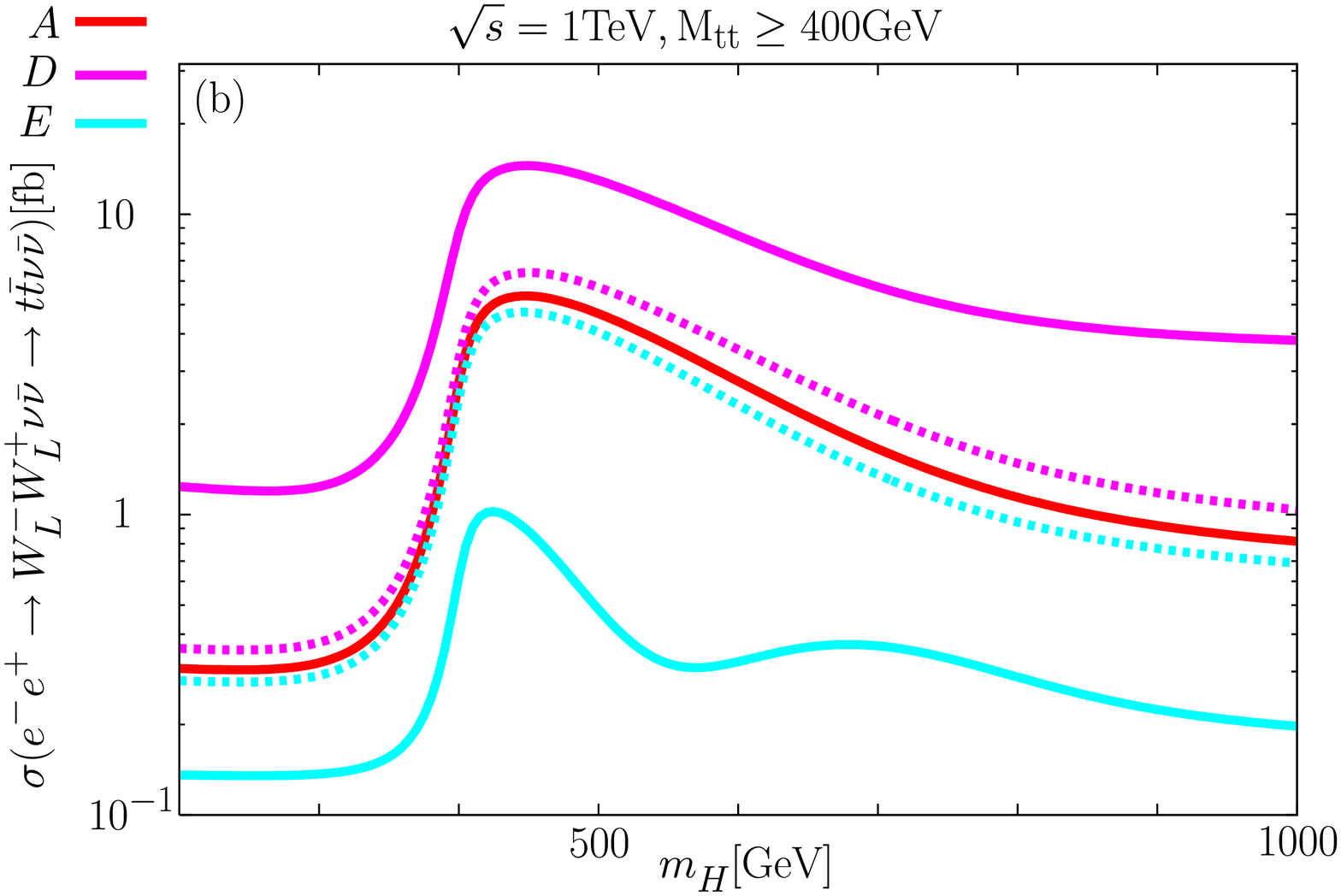}}
\end{minipage}
\caption{Cross sections of
 $e^-e^+\to W^-_L W^+_L \nu \bar \nu \to t \bar t \nu \bar \nu$
 are shown as a function of Higgs boson mass for the cases of
 Set A, Set B and Set C [Fig.~\ref{fig:SigmaCUT400-t1}(a)], and
 for those of Set A, Set D and Set E [Fig.~\ref{fig:SigmaCUT400-t1}(b)]
 with $\Lambda=1$ TeV (solid curves) and $3$ TeV (dashed curves).
 The collider energy is taken to be $\sqrt{s}=1$ TeV.
}
\label{fig:SigmaCUT400-t1}
\end{figure}

\begin{figure}[t]
\begin{minipage}{8cm}
\unitlength=1cm
\rotatebox{0}{
\includegraphics[width=7.5cm]{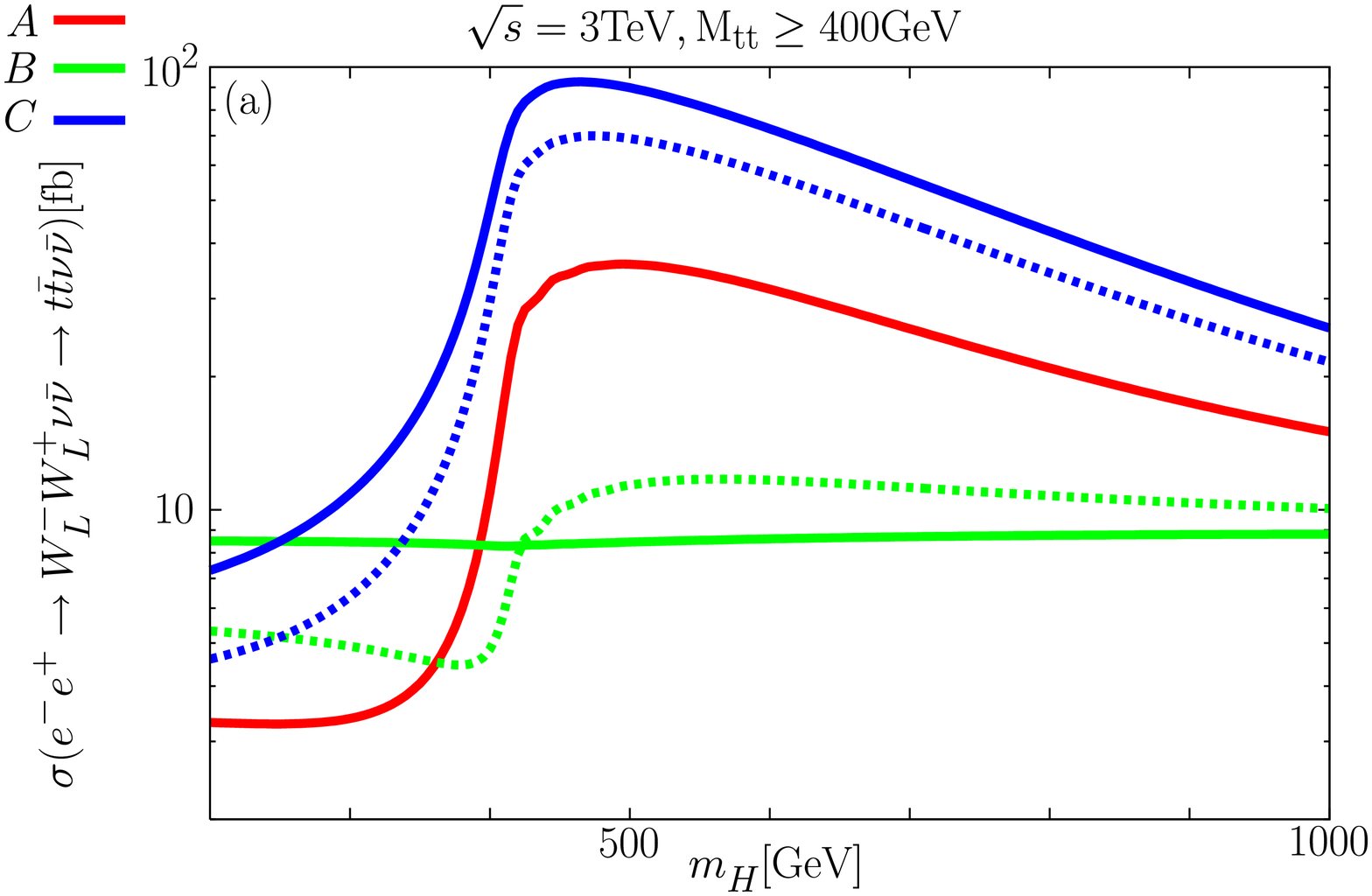}}
\end{minipage}
\begin{minipage}{8cm}
\unitlength=1cm
\rotatebox{0}{
\includegraphics[width=7.5cm]{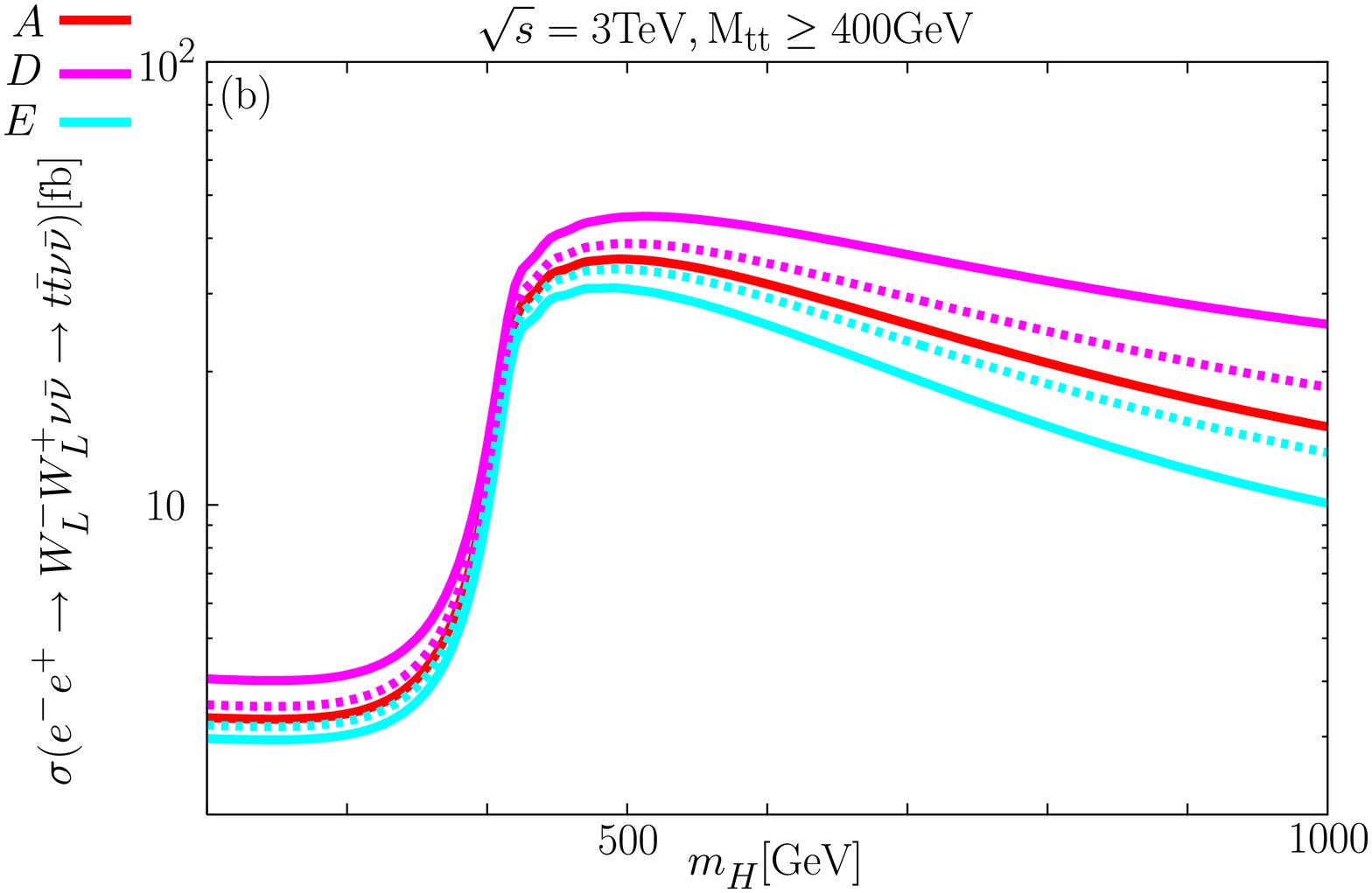}}
\end{minipage}
\caption{Cross sections of 
 $e^-e^+\to W^-_L W^+_L \nu \bar \nu \to t \bar t \nu \bar \nu$
 are shown as a function of Higgs boson mass for the cases of
 Set A, Set B and Set C [Fig.~\ref{fig:SigmaCUT400-Dt}(a)], and those of 
 Set A, Set D and Set E [Fig.~\ref{fig:SigmaCUT400-Dt}(b)]
 with $\Lambda=3$ TeV (solid curves) and $5$ TeV (dashed curves).
 The collider energy is taken to be $\sqrt{s}=3$ TeV.
}
\label{fig:SigmaCUT400-Dt}
\end{figure}

Now we turn to the discussion on the case with the dimension-six operators
${\cal O}_{t1}$ and ${\cal O}_{Dt}$. 
In Figs.~\ref{fig:SigmaCUT400-t1}(a)  and \ref{fig:SigmaCUT400-t1}(b)  
  cross sections for $e^- e^+ \to W^-_L W^+_L
 \nu \bar \nu \to  t \bar{t} \nu \bar{\nu}$ 
 are shown as a function of $m_H^{}$
 with the kinematic cut $M_{tt} \ge 400$ GeV.
 The collider energy is set to be $\sqrt{s}=1$ TeV.
 The new physics scale $\Lambda$ is assumed to be 1 TeV and 3 TeV.
 Fig.~\ref{fig:SigmaCUT400-t1}(a) shows 
 the results for Set B and Set C, and  
 Fig.~\ref{fig:SigmaCUT400-t1}(b) does those for Set D and Set E. 
 In the both figures, the result in the SM case [Set A] is also
 plotted. 

 The cases for a higher collider energy ($\sqrt{s}=3$ TeV) are shown in 
 Figs.~\ref{fig:SigmaCUT400-Dt}(a)  and \ref{fig:SigmaCUT400-Dt}(b).
 The new physics scale $\Lambda$ is assumed to be 3 TeV and 5 TeV.
 Fig.~\ref{fig:SigmaCUT400-Dt}(a) shows 
 the results for Set B and Set C, and  
 Fig.~\ref{fig:SigmaCUT400-Dt}(b) does those for Set D and Set E. 
 In the both figures, the result in the SM case [Set A] is also
 plotted.
 The cross sections for $\sqrt{s}=3$ TeV
 in Figs.~\ref{fig:SigmaCUT400-Dt}(a) and (b)
 are greater than that for $\sqrt{s}=1$ TeV
 in Figs.~\ref{fig:SigmaCUT400-t1}(a) and (b) due to logarithmic
 collinear enhancement at the $W$ bosons in the fusion process.
 The deviation rates from the SM prediction for Set B and Set C
 (the effect of $C_{t1}$) in Fig.~\ref{fig:SigmaCUT400-Dt}(a)
 can be huge and is similar to those in
 Fig.~\ref{fig:SigmaCUT400-t1}(a),
 while those for Set D and Set E (the effect of $C_{Dt}$)
 in Fig.~\ref{fig:SigmaCUT400-Dt}(b) are much smaller than
 those in Fig.~\ref{fig:SigmaCUT400-t1}(b).   
 These differences in the energy dependence between the effects
 of $C_{t1}$ and $C_{Dt}$ may be used to extract their contributions 
 separately at a multi-TeV collider such as CLIC. 
 
\begin{figure}[t]
\unitlength=1cm
\rotatebox{0}{
\includegraphics[width=9cm]{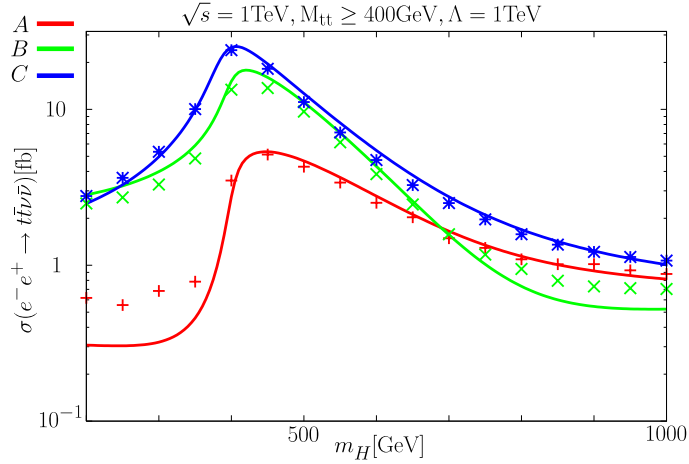}}
\caption{
 Cross sections for $e^- e^+ \to t\bar t \nu \bar \nu$
 evaluated in the full matrix-element calculation (marked points)
 for the cases of Set A, Set B and Set C. The other parameters are taken
 to be $m_H^{} = 500$ GeV, $\sqrt{s} = 1$ TeV and $\Lambda = 1$ TeV.
 The corresponding results for
 $e^- e^+ \to W^{+}_LW^{-}_L\nu\bar\nu\to t\bar t \nu \bar \nu$
 by using the EWA are also plotted 
 for comparison (solid curves).
}
\label{fig:ewa-vs-calchep}
\end{figure}

In Fig.~\ref{fig:ewa-vs-calchep}, we show the consistency
of the numerical evaluation of the cross section.
We plot the points evaluated by the full matrix-element
calculation of the cross section for
$e^- e^+ \to 
t \bar t \nu \bar \nu$ in CalcHEP\cite{calchep} and LanHEP\cite{lanhep}
for the cases of Set A, Set B and Set C. 
The solid curves show the results of cross sections for $e^-
e^+ \to W^-_L W^+_L\nu\bar\nu\to t \bar t \nu \bar \nu$
by using EWA for Set A, Set B and Set C.
Both results agree for greater Higgs boson mass. 
The both results agree with each other
in about 20-30 \% error for $m_H^{} \gsim 400$ GeV where
the longitudinal component of the $W$ bosons is dominant.
It is also found that even for relatively lower $m_H^{}$ values
($\lsim 400$ GeV) the deviation from the SM prediction (Set A)
evaluated by using CalcHEP/LanHEP are huge (more than 100 \%)
for both Set B and Set C. 
Therefore our conclusion originally obtained
from the analysis in the EWA can also be applied
to the small $m_H^{}$ region
where the EWA cannot be used for a valid calculation.
   For a light Higgs boson with its mass to be less than about 150 GeV,
   the Higgs-top interaction is expected to be
   studied via the process  $e^-e^+ \to t \bar t H$\cite{han}.  

 The SM value of the cross section for $e^-e^+ \to W^-W^+ \nu \bar \nu
 \to t \bar t \nu \bar \nu$ is order 1fb for heavy Higgs bosons
 ($m_{H}^{} \gsim 400$ GeV).  
 At the ILC with an $e^-e^+$ energy of 1 TeV and
 the integrated luminosity of 500fb$^{-1}$, over several times
 100 of the events are produced. 
 Naively, the statistic error of the cross section measurement
 can be less than  about 10\% level.
 The QCD corrections are evaluated to be the same order of magnitude\cite{godfrey}.
 Therefore, we can expect that the large enhancement of the
 cross section due to the contribution of
 ${\cal O}_{t1}$ can be easily observed as long as it changes
 the cross section by a few times 10\% or more.   
 The effect of ${\cal O}_{Dt}$ under the constraint from LEP results
 may also be observed as long as it changes the SM cross section by
 10-20 \%.
 
 The backgrounds should also be taken into account.
 One of the main backgrounds is
 $e^-e^+\to \gamma t\bar t$ with $\gamma$ to be missing.
 It is known that a kinematic cut for the transverse momentum of
 the final top quark can reduce this mode significantly. 
 In Ref.~\cite{larios}, the simulation study for the background reduction
 has been done in the SM, and it is concluded that
 the background can be sufficiently suppressed by the kinematic cuts.
 Another important background is the top pair production process
 via the photon fusion $\gamma\gamma\to t \bar t$.
 This mode can be suppressed by the cut $E\!\!\!\!/ > 50$
 GeV\cite{alcaraz}, where $E\!\!\!\!/$ is the missing energy.
 Finally, the direct top-pair production $e^-e^+ \to t \bar t$
 can easily be suppressed by imposing the cut for the invariant
 mass $M_{tt}$.

   We have mainly shown the results with the possible maximal values
   of $C_{t1}$ and $C_{Dt}$.
   For smaller values of these couplings, the effect
   becomes reduced according to the size of the couplings.
   In such a case, the helicity analysis of the final
   top (anti-top) quark would make it possible 
   to discriminate the contribution from the
   Higgs boson mediated diagram.
   The systematic analysis along this line will be performed
   for these intermediate regions of the Higgs boson mass
   in our future publications\cite{kmt}.
 
 Finally, in this paper we have concentrated on the process at $e^-e^+$
 colliders, and we have found that a deviation from the
 SM result can be a few  times 100 \% 
 greater than the SM prediction, which can easily be detectable
 at a linear collider. 
 Such a large deviation may also be detectable at
 hadron colliders via the process such as
 $pp \to W^-W^+ X\to t \bar t X$ even though the
 QCD background is huge.
 We may also consider a similar process such as  
 $pp \to W^-W^+ X\to \tau \bar \tau X$.
 In any case, detailed simulation study should be necessary
 to clarify the significance.

\section{Conclusions}
\label{Sec:conclusion}

 In the low energy effective theory,
 information of the dynamics of mass generation is described
 by  higher dimensional operators which affect the SM top Yukawa
 coupling.
 We have discussed the effect of non-standard interactions
 characterized by dimension-six operators on the effective top
 Yukawa coupling.
 We have evaluated the effects of the dimension-six operators
 ${\cal O}_{t1}$ and ${\cal O}_{Dt}$ on the cross section for the 
 process $e^-e^+ \to W^-W^+\nu \bar \nu \to t \bar t \nu \bar \nu$.
 The magnitude of the coefficients of these dimension-six operators
 are constrained by perturbative unitarity and current experimental
 data. We then studied possible deviation from the SM prediction
 in these cross sections.
 We have found that the effect of the dimension-six operator ${\cal O}_{t1}$
 can enhance the cross section by a few times 100\% for heavy
 Higgs bosons ($m_h > 2 m_t$) when the new physics scale $\Lambda$ is in
 a TeV region.
 Such a large deviation from the SM prediction
 should be detectable at the ILC and the CLIC.
 The detailed study for a Higgs boson with an intermediate mass 
 ($150 {\rm GeV} < m_H^{} < 2 m_t$) by including helicity analysis for
 the final top quarks will be presented elsewhere.  
 
\vspace{1cm}
\noindent
{\large \it Acknowledgments}\\
The authors would like to thank
Bohdan Grzadkowski,
Kaoru Hagiwara,  
Zenro Hioki,
Koichi Matsuda,  
Kazumasa Ohkuma,
and C.-P. Yuan for valuable discussions. 
S.K. was supported, in part, by Grants-in-Aid of the Ministry 
of Education, Culture, Sports, Science and Technology, Government of 
Japan, Grant Nos. 17043008 and 18034004.

\newpage
\appendix

\noindent
 \section*{Appendix}

The dimension six operators ${\cal O}_{t1}$ and ${\cal O}_{Dt}$
are defined in Eqs.~(\ref{eq:Ot1}) and (\ref{eq:ODt}).
We here list the explicit expressions of all the terms in the r.h.s.
of Eqs.~(\ref{eq:dec-Ot1}) and (\ref{eq:dec-ODt}) in terms of
the component fields.

\subsection{The operator $O_{t1}^{}$}
 
\begin{itemize}
\item The 3-vertex \\
\begin{align}
\mathcal O_{t1}^{v^2\phi \bar \psi \psi} = \frac{v^2}{\sqrt2} H \bar t t.
\end{align}
\item The 4-vertex \\
\begin{align}
\mathcal O_{t1}^{v\phi^2 \bar \psi \psi} = &\frac{3v}{2\sqrt2} H^2 \bar t t 
+ \frac{v}{2\sqrt2} z^2 \bar t t + \frac{v}{\sqrt2} \omega^+ \omega^- \bar t t 
\nonumber \\
&-i v H z \left( \bar t_L^{} t_R^{} - \bar t_R^{} t_L^{} \right) 
-v H \left( \omega^- \bar b_L^{} t_R^{} + \omega^+ \bar t_R^{} b_L^{} \right).
\end{align}
\item The 5-vertex \\
\begin{align}
\mathcal O_{t1}^{\phi^3 \bar \psi \psi} = &\frac{1}{2\sqrt2} H^3 \bar t t + \frac{1}{2\sqrt2} H z^2 \bar t t + \frac{1}{\sqrt2} H \omega^- \omega^+ \bar t t \nonumber \\
&-\frac{i}{2\sqrt2} H^2 z \left( \bar t_L^{} t_R^{} - \bar t_R^{} t_L^{} \right) -\frac{i}{2\sqrt2} z^3 \left( \bar t_L^{} t_R^{} - \bar t_R^{} t_L^{} \right) -\frac{i}{\sqrt2} z \omega^+ \omega^- \left( \bar t_L^{} t_R^{} - \bar t_R^{} t_L^{} \right) \nonumber \\
&-\frac{1}{2} H^2 \left( \omega^- \bar b_L^{} t_R^{} + \omega^+ \bar
 t_R^{} b_L^{} \right) -\frac{1}{2} z^2 \left( \omega^- \bar b_L^{}
 t_R^{} + \omega^+ \bar t_R^{} b_L^{} \right) \nonumber\\
& - \omega^+ \omega^- \left( \omega^- \bar b_L^{} t_R^{} + \omega^+ \bar t_R^{} b_L^{} \right).
\end{align}
\end{itemize} 

\begin{itemize}
\subsection{The operator $O_{Dt}^{}$}
\item The 3-vertices \\
\begin{align}
\mathcal O_{Dt}^{\partial \phi \bar \psi \partial \psi} = 
&\frac{1}{\sqrt2} \partial_\mu H \left( \bar t_L^{} \partial^\mu t_R^{}
 + \partial^\mu \bar t_R^{} t_L^{} \right) - \frac{i}{\sqrt2}
 \partial_\mu z \left( \bar t_L^{} \partial^\mu t_R^{} - \partial^\mu
 \bar t_R^{} t_L^{} \right)\nonumber\\
&- \partial_\mu \omega^- \bar b_L^{} \partial^\mu t_R^{} - \partial_\mu \omega^+ \partial^\mu \bar t_R^{} b_L^{}. 
\end{align}
\begin{align}
\mathcal O_{Dt}^{v V \bar \psi \partial \psi} = 
&-i\frac{g_Z^{}v}{2\sqrt2} Z^\mu \left( \bar t_L^{} \partial^\mu t_R^{} - \partial^\mu \bar t_R^{} t_L^{} \right) -i\frac{gv}{2} \left( {W^-}_\mu \bar b_L^{} \partial^\mu t_R^{} - {W^+}_\mu \partial^\mu \bar t_R^{} b_L^{} \right).
\end{align}
\item The 4-vertices \\
\begin{align}
\mathcal O_{Dt}^{v V^2 \bar \psi  \psi} = 
&-\frac{g_Z^{}v}{2\sqrt2} Q_t e Z_\mu A^\mu \bar t t + \frac{g_Z^2 v}{2\sqrt2} s_W^2 Q_t Z_\mu Z^\mu \bar t t \nonumber \\ 
&- \frac{gv}{2} Q_t e A_\mu \left( {W^-}^\mu \bar b_L^{} t_R^{}
+ {W^+}^\mu \bar t_R^{} b_L^{} \right)  \nonumber\\
 &- \frac{gv}{2} g_Z^{} s_W^2 Q_t Z_\mu \left( {W^-}^\mu \bar b_L^{} t_R^{} + {W^+}^\mu \bar t_R^{} b_L^{} \right). 
\end{align}
\begin{align}
\mathcal O_{Dt}^{V \phi \bar \psi \partial \psi} = 
&-i\frac{g_Z^{}}{2\sqrt2} Z_\mu H \left( \bar t_L^{} \partial^\mu t_R^{} - \partial^\mu \bar t_R^{} t_L^{} \right) - \frac{g_Z^{}}{2\sqrt2} Z_\mu z \left( \bar t_L^{} \partial^\mu t_R^{} + \partial^\mu \bar t_R^{} t_L^{} \right) \nonumber \\
&-i\frac{g}{\sqrt2} \left( {W^+}_\mu \omega^+ \bar t_L^{} \partial^\mu t_R^{} - {W^-}_\mu \omega^+ \partial^\mu \bar t_R^{} t_L^{} \right) \nonumber \\
&-i e A_\mu \left( \omega^- \bar b_L^{} \partial^\mu t_R^{} - \omega^+ \partial^\mu \bar t_R^{} b_L^{} \right) -i \frac{g_Z^{}}{2} \left( 1-2 s_W^2 \right) Z_\mu \left( \omega^- \bar b_L^{} \partial^\mu t_R^{} - \omega^+ \partial^\mu \bar t_R^{} b_L^{} \right) \nonumber \\
&-i \frac{g}{2} H \left( {W^-}_\mu \bar b_L^{} \partial^\mu t_R^{} - {W^+}_\mu \partial^\mu \bar t_R^{} b_L^{} \right) - \frac{g}{2} z \left( {W^-}_\mu \bar b_L^{} \partial^\mu t_R^{} + {W^+}_\mu \partial^\mu \bar t_R^{} b_L^{} \right).
\end{align}
\begin{align}
\mathcal O_{Dt}^{V \partial \phi \bar \psi \psi} = 
&-\frac{1}{\sqrt2} Q_t e A_\mu \partial^\mu z \bar t t + \frac{1}{\sqrt2} g_Z^{} s_W^2 Q_t Z_\mu \partial^\mu H \bar t t \nonumber \\
&-\frac{i}{\sqrt2} Q_t e A_\mu \partial^\mu H \left( \bar t_L^{} t_R^{} - \bar t_R^{} t_L^{} \right) + \frac{i}{\sqrt2} g_Z^{} s_W^2 Q_t Z_\mu \partial^\mu H \left( \bar t_L^{} t_R^{} - \bar t_R^{} t_L^{} \right) \nonumber \\
& + i Q_t e A_\mu (\partial^\mu \omega^- \bar b_L^{} t_R^{} - \partial_\mu \omega^+ \bar t_R^{} b_L^{}) -i g_Z^{} s_W^2 Q_t Z_\mu (\partial^\mu \omega^- \bar b_L^{} t_R^{} - \partial_\mu \omega^+ \bar t_R^{} b_L^{}).
\end{align}
\item The 5-vertices \\
\begin{align}
\mathcal O_{Dt}^{V^2 \phi \bar \psi \psi} = 
& -\frac{g_Z^{}}{2\sqrt2} Q_t e A_\mu Z^\mu H \bar t t + \frac{g_Z^2}{2\sqrt2} s_W^2 Q_t Z_\mu Z^\mu H \bar t t \nonumber \\ 
&+i \frac{g_Z^{}}{2\sqrt2} Q_t e A_\mu Z^\mu z \left( \bar t_L^{} t_R^{} - \bar t_R^{} t_L^{} \right) -i \frac{g_Z^2}{2\sqrt2} s_W^2 Q_t e Z_\mu Z^\mu z \left( \bar t_L^{} t_R^{} - \bar t_R^{} t_L^{} \right) \nonumber \\ 
&+\frac{g}{\sqrt2} Q_t e A_\mu \left( {W^+}^\mu \omega^- \bar t_L^{}
 t_R^{} + {W^-}^\mu \omega^+ \bar t_R^{} t_L^{} \right)\nonumber\\
& - \frac{g}{\sqrt2} g_Z^{} s_W^2 Q_t Z_\mu \left( {W^+}^\mu \omega^- \bar t_L^{} t_R^{} + {W^-}^\mu \omega^+ \bar t_R^{} t_L^{} \right) \nonumber \\
&- Q_t e^2 A_\mu A^\mu \left( \omega^- \bar b_L^{} t_R^{} + \omega^+ \bar t_R^{} b_L^{} \right) - g_Z^{} \left( \frac12 - 2 s_W^2 \right) Q_t e A_\mu Z^\mu \left( \omega^- \bar b_L^{} t_R^{} + \omega^+ \bar t_R^{} b_L^{} \right) \nonumber \\ 
&+ \frac{g_Z^2}{2} \left( 1 - 2 s_W^2 \right) s_W^2 Q_t Z_\mu Z^\mu \left( \omega^- \bar b_L^{} t_R^{} + \omega^+ \bar t_R^{} b_L^{} \right) \nonumber \\
&-\frac{g}{2} Q_t e A_\mu H \left( {W^-}^\mu \bar b_L^{} t_R^{} + {W^+}^\mu \bar t_R^{} b_L^{} \right) +\frac{g}{2} g_Z^{} s_W^2 Q_t Z_\mu H \left( {W^-}^\mu \bar b_L^{} t_R^{} + {W^+}^\mu \bar t_R^{} b_L^{} \right) \nonumber \\
&+i\frac{g}{2} Q_t e A_\mu z \left( {W^-}^\mu \bar b_L^{} t_R^{} -
 {W^+}^\mu \bar t_R^{} b_L^{} \right)\nonumber\\
& -i\frac{g}{2} g_Z^{} s_W^2 Q_t Z_\mu z \left( {W^-}^\mu \bar b_L^{} t_R^{} - {W^+}^\mu \bar t_R^{} b_L^{} \right).
\end{align}
\end{itemize}

\end{document}